\documentclass[iop]{emulateapj}
\pdfoutput=1
\usepackage{amsmath}
\usepackage{amsfonts}
\usepackage{amssymb}
\usepackage{natbib}
\usepackage{hyperref}
\hypersetup{pdfborder={0 0 0}, colorlinks=true, citecolor=blue, pdftitle=Towards a Realistic Global Corona Model}

\submitted{submitted 12/08/2009 to ApJ}

\shorttitle{Towards a Realistic Global Corona Model}
\shortauthors{Downs et al.}

\begin{document}

\title{Towards a Realistic, Data-Driven Thermodynamic MHD Model of the Global Solar Corona}

\author{Cooper Downs\altaffilmark{1}, Ilia I. Roussev\altaffilmark{1},  Bart van der Holst\altaffilmark{2}, No\'e Lugaz\altaffilmark{1}, Igor V. Sokolov\altaffilmark{2}, and Tamas I. Gombosi\altaffilmark{2}}
\email{cdowns@ifa.hawaii.edu}
\altaffiltext{1}{Institute for Astronomy, University of Hawaii, at Manoa, 2680 Woodlawn Dr., Honolulu, HI 96822, USA}
\altaffiltext{2}{Center for Space Environment Modeling, University of Michigan, 2455 Hayward St., Ann Arbor, MI 48109, USA}

\begin{abstract}

In this work we describe our implementation of a thermodynamic energy equation into the global corona model of the Space Weather Modeling Framework (SWMF), and its development into the new Lower Corona (LC) model. This work includes the integration of the additional energy transport terms of coronal heating, electron heat conduction, and optically thin radiative cooling into the governing magnetohydrodynamic (MHD) energy equation. We examine two different boundary conditions using this model; one set in the upper transition region (the Radiative Energy Balance model), as well as a uniform chromospheric condition where the transition region can be modeled in its entirety.  Via observation synthesis from model results and the subsequent comparison to full sun extreme ultraviolet (EUV) and soft X-Ray observations of Carrington Rotation (CR) 1913 centered on Aug 27, 1996, we demonstrate the need for these additional considerations when using global MHD models to describe the unique conditions in the low corona.  Through multiple simulations we examine ability of the LC model to asses and discriminate between coronal heating models, and find that a relative simple empirical heating model is adequate in reproducing structures observed in the low corona. We show that the interplay between coronal heating and electron heat conduction provides significant feedback onto the 3D magnetic topology in the low corona as compared to a potential field extrapolation, and that this feedback is largely dependent on the amount of mechanical energy introduced into the corona.

\end{abstract}


\section{Introduction and Background}

While the inherent complexity of the solar corona and its dynamic processes has engendered great scientific interest and study over the past half century, it is this complexity that is perhaps the greatest challenge when attempting to model and understand the underlying processes of observed phenomena in a quantitative fashion \citep{aschwanden08:asp}. For this reason, the development and use of fully 3D magnetohydrodynamic (MHD) models describing the corona and solar wind, which has undergone a veritable renaissance in the computing age, has been critical to furthering our understanding of the solar wind and eruptive events \citep{roussev06}. A particularly exciting avenue has been the development of 3D observation-based techniques, which provide the opportunity to include realistic conditions of the corona into studies of the global solar wind \citep{mikic99,roussev03b,cohen07} and to address how this governs dynamic events \citep{lugaz07,roussev07}. 
\\

With its origin in the long-studied fundamental question of how the solar corona can be heated to and maintain temperatures in the million Kelvin (MK) regime, a primary focus of this study is to address the issue of energy input into the corona in the context of a fully 3D global model.  Because non-MHD processes of energy transport and local conditions in the low corona are responsible for coronal heating mechanisms, it is often quite difficult and computationally expensive to include the small-scale micro physics of reconnection and turbulence responsible for coronal heating in 3D models, especially since the exact processes involved are not universally agreed upon.  As a result, heating models are often parameterized as a heating term that depends on various local magnetic and thermodynamic properties that is included in the energy equation (e.g. \citet{aschwanden02,schrijver04,abbett07,mok08}).  However, due to the large number of heating models and their ad-hoc formulations, it is crucial to use real observations to compare and constrain these models and deduce the best combination of models, both for active regions and the quiet Sun, that yields the most realistic global simulations. When using such models to study transient events in the Solar Corona, particularly Coronal Mass Ejections (CMEs) and coronal waves seen in the extreme ultraviolet regime (EUV waves, \citet{moses97}), it is a motivating principle that only by constraining physical theories and scenarios through as many observable manifestations available can we provide the best possible avenue to further our knowledge and precision in understanding these dynamic events (e.g. \citet{Lugaz08,Lugaz09, cohen09}). The ultimate goal being to strip away layers of empirical approximation with more fundamental physical terms and mechanisms. However, it is equally important to validate and asses the ability of these improvements to adequately represent the global conditions in the solar corona.
\\
\clearpage
To address this issue through a deterministic, data-driven approach, we modify the global corona model of the Space Weather Modeling Framework (SWMF) \citep{toth05} to include the transition region between the chromosphere and the corona, where non-MHD thermodynamic terms of energy transport, such as electron heat conduction, radiative losses, and coronal heating all become important.  Using techniques similar to the pioneering work of \citet{lionello01,lionello09}, in this refined version of the model the inner boundary is placed at the chromosphere or upper transition region rather than in the low corona and the now relevant non-MHD terms are added to the governing MHD energy equation.
\\

In section \ref{section:Model} we discuss the new Lower Corona (LC) model and physical considerations added as part of this work. In section \ref{section:Model_Runs} we overview our runs and results and discuss the application of this model to the study of the corona (in particular the comparison to EUV imaging observations). In section \ref{section:Detailed_Analysis} we provide a detailed analysis of a particular run, and benchmark it via comparison to a previous corona model. We conclude in section \ref{section:conclusion}.

\section{The Simulation Tool}

\label{section:Model}

\subsection{SWMF}The main model and starting point for this work is the SWMF Solar Corona (SC) model \citep{cohen07}. As far as numerical methods, the code is fully parallelized and designed to be highly customizable in terms the methods, solvers, and equations used. A complete description of the MHD equations and their implementation in the SWMF can be found in \citet{powell99} as well as the addition of source terms in \citep{groth00}. For the current version of the Solar Corona Model, a steady state solar wind solution is obtained using a variable $\gamma$ model \citep{roussev03b,cohen07,cohen08}. The latter references use the Wang-Sheeley-Arge (WSA) model \citep{arge04} as an initial condition to derive the variation in the polytropic index. This then achieves a steady state MHD solution after nominal integration (henceforth refered to as the ``Standard SC model'').
\\ 
\label{section:Model_swmf}

Physically, the most important advantage of this tool and others like it lies in its ability to simulate the complete 3D environment of any event and Carrington Rotation (standardized solar rotation number, abbreviated CR) for which data for the photospheric magnetic field is available throughout the entire solar surface. The initial magnetic configuration is extrapolated using the Potential Field Source Surface Method (PFSSM) \citep{altschuler77}, which uses magnetic coefficients derived from observations of the synoptic photospheric magnetic field (typically high order maps using data from the MDI instrument aboard the SOHO observatory, or low order maps the Wilcox Solar Observatory). Using this method, the subsequent evolution of the magnetic field towards a steady state during the simulation is no longer strictly potential.
\\

While the standard version of the SC model has been very successful at reproducing the bi-modal solar wind structure and observations at 1AU \citep{cohen08}, which was the fundamental goal of that empirical model, the lower boundary at the solar surface is quite smooth and nearly uniform. In order to reproduce and study the fine structures of the the low corona we significantly modify the SC model to address the unique physics that take place at this boundary (henceforth refered to as the the Lower Corona (LC) model).

\subsection{Including Additional Thermodynamic Terms}This modification of the energy equation takes the form of:
\begin{equation} \label{newterms}
\frac{\partial{E}}{\partial{t}} + \nabla\cdot(\vec{F}_{MHD} + \vec{F}_c) = Q_{MHD} + Q_{r} + Q_{h}
\end{equation}
where, the subscript $MHD$ refers to the standard MHD terms in the SWMF and the additional terms $\vec{F}_c$, $Q_{r}$, $Q_{h}$ are described below. Because these terms account for realistic energy transport, the polytropic index is no longer variable, and is set to a uniform value of $\gamma=5/3$.

\subsubsection{Heat Conduction}The standard form of anisotropic electron heat conduction in the  collisional limit is included as an additional energy flux term, $\vec{F_c} = -\kappa_0 T^{5/2}\hat{B}(\hat{B}\cdot\nabla{T})$, in the energy equation (equivalent physically to adding a source term of $Q_c = \nabla \cdot \vec{F_c})$. The parallel component of Spitzer conductivity \citep{spitzer65}, this term is especially important in the transition region and low corona and plays a critical role in determining the equilibrium density and temperatures at the base of the corona. We use a value of $\kappa_0 = 1.23 \times 10^{-6}$ in cgs units.
\\

We also include a method for broadening the equilibrium scale length of the transition region, which is needed to resolve the entire transition region when using a chromospheric boundary (described in detail by \citet{abbett07,lionello09}). This reduces to both fixing $\kappa_0 T^{5/2}$ to a constant value of $\kappa_0 T_{mod}^{5/2}$ and reducing the radiative cooling term by the inverse factor $(T/T_{mod}^{5/2}$) for temperatures below a transition region value, $T<T_{mod}$. In this work we use $T_{mod} = 300,000\text{K}$.
\\

\label{section:heat_conduction}

\subsubsection{Radiative Losses}The next term, $Q_r = -n_en_p\Lambda(T)$, accounts for the radiative losses of the hot coronal plasma in the optically thin limit. The loss function $\Lambda(T)$ is calculated using the CHIANTI version 5 radiative loss routines \citep{landi06} and linearly interpolated as a tabulated function in the model. The assumption of a fully ionized hydrogen gas is already implicit in this model, which gives $n_en_p \sim n_e^2$. It is important to note that the choice of abundances used in calculating the radiative losses function can change the total losses significantly in the coronal temperature regime. For this reason we choose to use the coronal abundances file: \texttt{sun\_coronal\_ext.abund} \citep{landi02} when calculating the radiative cooling curve with CHIANTI, and do not vary this function in this work.

\clearpage

\subsubsection{Heating Model 1 - $|B|$ weighted}The first empirical coronal heating term, $Q_h$, that we examine is equivalent in form to that used by \citet{abbett07} and was originally based on work relating the surface unsigned magnetic flux of a convective envelope to the x-ray luminosity of stars \citep{bercik05}. This model is expressed as:
\begin{equation} \label{qheat}
Q_h = \frac{c\phi^\alpha \psi}{\zeta\int \psi \ dV}.
\end{equation}

\label{heating_model_1}
Where $c=0.8940$, $\alpha=1.1488$ are fixed parameters, $\phi$ represents the total unsigned magnetic flux at the solar surface, $\psi$ is the local heating weighting function, and $\zeta$ is a normalization constant. This formulation calculates the total amount of energy input into the corona and distributes it via $\psi$, which is chosen to be a function of $|B|$. The primary benefit of this model is its simplicity, the amount of heating at any location is given by the value of $\psi$, and the total amount of heat input to the corona (nearly constant) is normalized by the global integral in the denominator, $\int \psi \ dV$, which is carried out over the entire domain. In this implementation, $\phi$ is calculated by integrating the unsigned flux at the model boundary, $\phi=\oint_{R=R_{Sun}}{|B_r|dS}$, which is essentially the photospheric magnetogram used to initiate the model.
\\ 

Because both the relationship between X-Ray Luminosity and total power input into the corona is not well constrained, and varying magnetogram resolution does not constrain the true unsigned flux, $\zeta$ represents a relatively free parameter with which to adjust the total power on a case by case basis. In our model runs we choose $\zeta = 1/50$ and linear weighting with the magnetic field magnitude,  $\psi(|B|) = |B|$. This gives the heating term the operational equivalent of $Q_h = H|B|$, where $H$ is typically around $\sim 4\times 10^{-5}$ in cgs units, and total integrated power of $Q_{tot} = 3.11\times 10^{28}\ \text{ergs s}^{-1}$ if applied uniformly.
\\

One drawback of this term (and of other $|B|$ weighted heating terms in global models) is that the heating scale height, defined as the point along a given loop where $Q_h(s) = Q_h(s=0)/e$, is not a free parameter and varies with the strength and spatial distribution of $\vec{B}$. Also, note that the value of $H$ depends on the specific global magnetic field configuration and the radius to which the volume integral is calculated. In this work, we limit the range of influence of this function to the low corona by multiplying $\psi$ by an exponential envelope function with a scale height of $40\text{Mm}$.

\subsubsection{Heating Model 2 - exponential heating} The second heating model is a simple exponential scale height model. Using a standard exponential form we define heating model 2 as:

\begin{equation} \label{qheat2}
Q_h = H_0 \exp{[-(r-R_\sun)/\lambda]}.
\end{equation}

\label{heating_model_2}
Where $H_0$ is the local heating rate at $r=\text{R}_\sun$ and $\lambda$ is the heating scale height. A short heating scale height is consistent with SOHO and TRACE EUV observations of coronal loops \citep{aschwanden00a,aschwanden01}, and motivated by the ``nanoflare'' class of coronal heating models, in which most of the heat is deposited in the upper chromosphere and transition region \citep{parker88}. It is important to note that it is the high thermal conductivity of the coronal plasma that subsequently distributes this energy along the field/flow lines. For heating quiet sun regions, we use the values $H_0 = 7.28\times 10^{-5} \ \text{ergs cm}^{-3}\ \text{s}^{-1}$, and $\lambda = 40\ \text{Mm}$, which give a total power of $Q_{tot} = 1.99\times 10^{28}\ \text{ergs s}^{-1}$ if applied uniformly (surface flux $= 3.27\times 10^{5} \ \text{ergs cm}^{-2}\ \text{s}^{-1}$). The lack of explicit dependence on $|B|$ with an exponential scale height model allows us examine directly the effect that purely the magnetic geometry of the 3D corona has on the resultant thermodynamic equilibrium.

\subsubsection{Coronal hole heating} To maintain realistic temperatures in the solar wind and coronal holes at large distances, we use the same form of eq. \eqref{qheat2} with values of $H_0 = 5.0\times 10^{-7} \ \text{ergs cm}^{-3}\ \text{s}^{-1}$, and $\lambda = 0.7 \text{R}_\sun$ giving additional total power $Q = 5.01\times 10^{27}\ \text{ergs s}^{-1}$. This function is applied uniformly in each model run. While not entirely realistic for the acceleration profiles observed in coronal hole regions and the solar wind, we find that including this term is adequate for reproducing coronal hole emission in the absence of short scale height heating.

\label{heating_model_CH}

\subsubsection{Open field cutoff} One interesting focus of this study is to isolate the complementary effects that non-uniform heating at the surface and the 3D magnetic topology  have in determining the equilibrium structure of the corona.  For this reason, as one of the options in the model we include the ability to apply the low scale height heating model (either 1 or 2) to only regions associated with closed field in the corona. To do so, we use the PFSSM extrapolation computed to initiate the model to determine at every computational element whether the field threading the location is open (connects to the source surface and becomes radial) or closed (turns back down to the solar surface). If the region is associated with open field, then only the long scale height ``coronal hole'' heating model is applied (section \ref{heating_model_CH}), while if the region is closed, both are applied. 
\\

We would also like to note that any of the empirical heating terms discussed above (sections \ref{heating_model_1} - \ref{heating_model_OC}) may also be attributed to wave absorption and dissapation, probably due to the ion interaction with Alfv\'en wave turbulence. In this case, the heating may be specified in terms of the wave (spectral) energy and their damping coefficient, and the wave turbulence may contribute not only to the energy source but also to the momentum fluxes (see e.g. \citet{sokolov09} and papers cited therein). However, this particular study is focused primarily on the plasma conditions low corona, and we operate under the assumption that Alfv\'en wave pressure will be much lower than the thermal pressure there.
\label{heating_model_OC}
\clearpage
\subsection{Boundary Conditions}
\label{section:boundary_conditions}
Boundary conditions at the $r=\text{R}_\sun$ surface are applied to fix constant values temperature and density at the face boundary according to either of the models described below. Each of these use a zero flow condition and fix the  tangential component of the magnetic field (similar to those used in \citet{roussev04,jacobs09}. For the supersonic flow at the outer boundary at $r=24\text{R}_\sun$ we use a floating (zero-gradient) condition.

\subsubsection{Chromospheric Boundary} The primary boundary condition used in this model is a simple uniform ``chromospheric'' condition. As described in \citet{lionello01,lionello09}, by setting chromospheric values of electron temperature, $T_e=2\times \text{10}^4\ \text{K}$, and density, $n_e=1\times \text{10}^{12}\ \text{cm}^{-3}$, and including the transition region broadening method described in section \ref{section:heat_conduction}, it becomes possible to resolve the entire transition region everywhere in a global model. This is a critical advantage because the resultant topological equilibrium in the coronal part of the solution (density enhancements, varying width of the transition region, etc.) will depend entirely on the included physics and magnetic geometry, and not on its proximity to the boundary of the model. This fact is particularly important for the study of any sort of dynamics in the global corona, as the sharp transition from fast Alfv\'en speed in the quiet corona ($V_a \sim 2-3 \times 10^2 \text{km/s}$) to extremely slow speeds in the chromosphere ($V_a \sim < 50 \text{km/s}$), provides a realistic dispersion buffer for fast magnetosonic waves.

\subsubsection{REB Model:}In order to include a realistic lower boundary condition for the corona, while at the same time maintaining decent computational efficiency for the global domain, we use the method of the Radiative Energy Balance Model (REB) outlined by \citet{withbroe88} (further formalism, including coronal heating in \citet{lionello01}). In short, this model fixes the base electron temperature to a high transition region value, $T^{tr}_e=5\times 10^5\ \text{K}$, and in turn assumes an equilibrium balance throughout the rest of the Transition Region below to the top of the Chromosphere, from which an electron density, $n^{tr}_e$, may be obtained. By rearranging the terms in the MHD energy equation and integrating over T at constant pressure, one obtains:
\begin{equation} \label{reb}
n^{tr}_e = \sqrt{\frac{\frac{1}{2}\kappa_0T_e^3(\hat{B}\cdot \nabla{T_e})^2 + \frac{2}{7}Q_hT_e^{\frac{3}{2}}}{\int_{T^{ch}_e}^{T^{tr}_e}T^{\frac{1}{2}}\Lambda(T)dT}}
\end{equation}
where, the integral of the radiative loss function is carried out between $T^{ch}_e=10^4$ and $T^{tr}_e=5\times 10^5\ \text{K}$. This allows for the base density $n^{tr}_e$ to be a function of the physical conditions included in the model (coronal heating, anisotropic heat conduction, and radiative cooling) and thus varies spatially along the boundary. Most importantly, the numerator of \eqref{reb} shows how the interplay between magnetic field strength, $|\vec{B}|$ (via $Q_h$ in the right term), and normal orientation at the surface, $\hat{B_r}$ (via $(\hat{B}\cdot \nabla{T_e})^2$ on the left), sets the base density on the solar surface. The ability to capture these features with such a simple, easily computed boundary is a primary motivation in using this particular model.

\subsection{Geometric Considerations}Another aspect of this work has been to address the unique geometric concerns that arise when including the transition region and below. A typical solar corona (SC) simulation is carried out with an  adaptive Cartesian grid over a Sun-centered $48\times 48\times  48$R$_{\sun}$ cube with $4\times4\times4$ cell blocks. The average cell size is smallest at the surface ($\sim 10^{-2}$ R$_{\sun} \sim 7,000$ km) and is incrementally increased with distance to a largest size of $0.65$ R$_{\sun}$ near the outer boundary, giving a few million cells. While this is a relatively large surface cell size, it is adequate considering the precision of the magnetic field boundary (an observational limitation). However, if one is required to resolve the transition region, with dynamic radial spatial scales on the order of a few km, it quickly becomes infeasible to use a Cartesian grid (with every $1/2$ refinement in length the number of surface cells increases by 4). This can be partially alleviated using methods to widen the extent of the model transition region to typical scales of order $300$ km without affecting the coronal solution (section \ref{section:heat_conduction}), but the problem is still quite substantial. This then represents a balance between height accuracy and computational efficiency (with a Cartesian grid, the number of surface cells increases by a factor of 4 with every successive $1/2$ refinement).\\
 
\begin{figure}[hbtp]
\centering
\hspace{0.5in}
\includegraphics[width=0.35\textwidth]{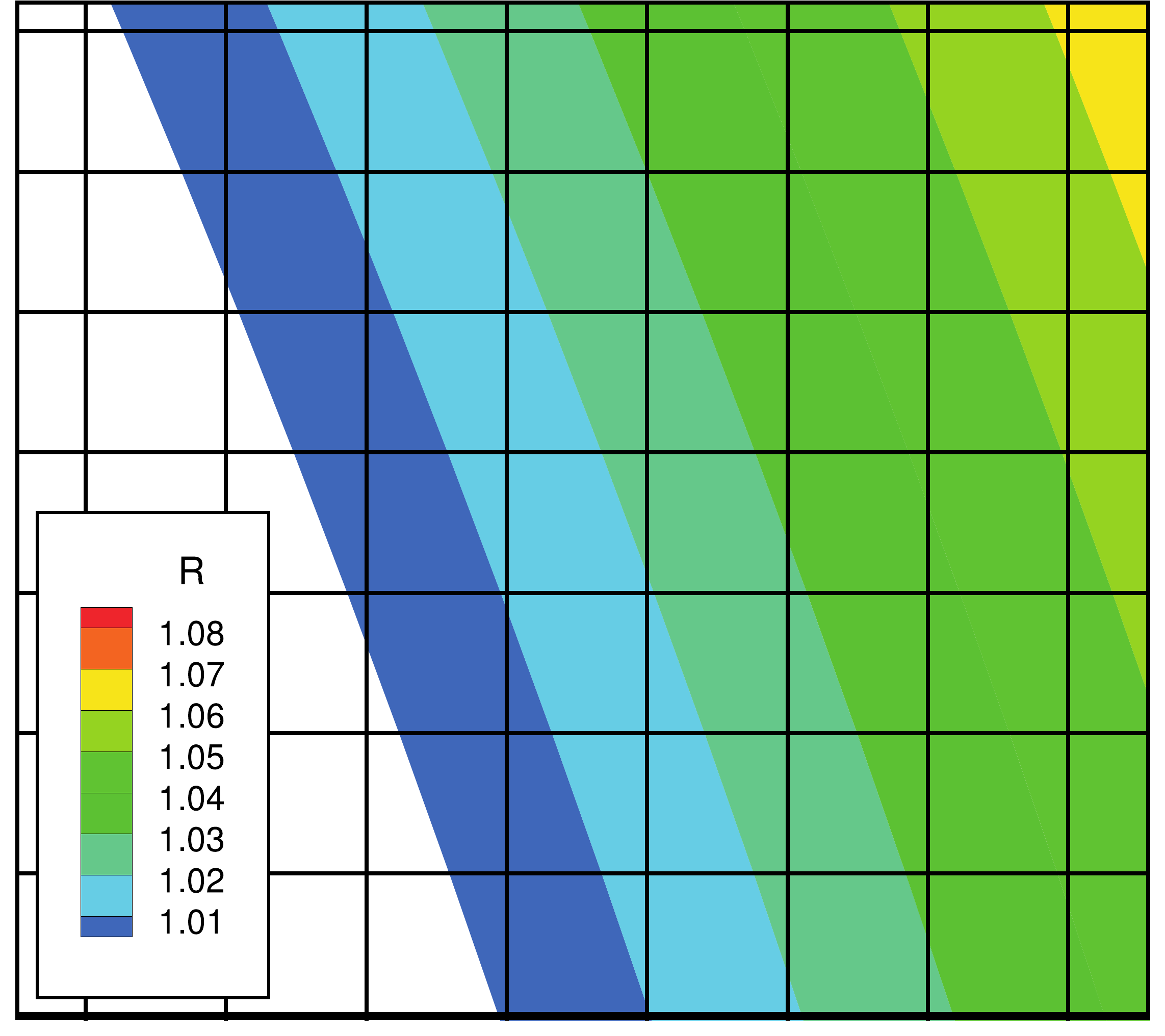}
\hfill
\includegraphics[width=0.35\textwidth]{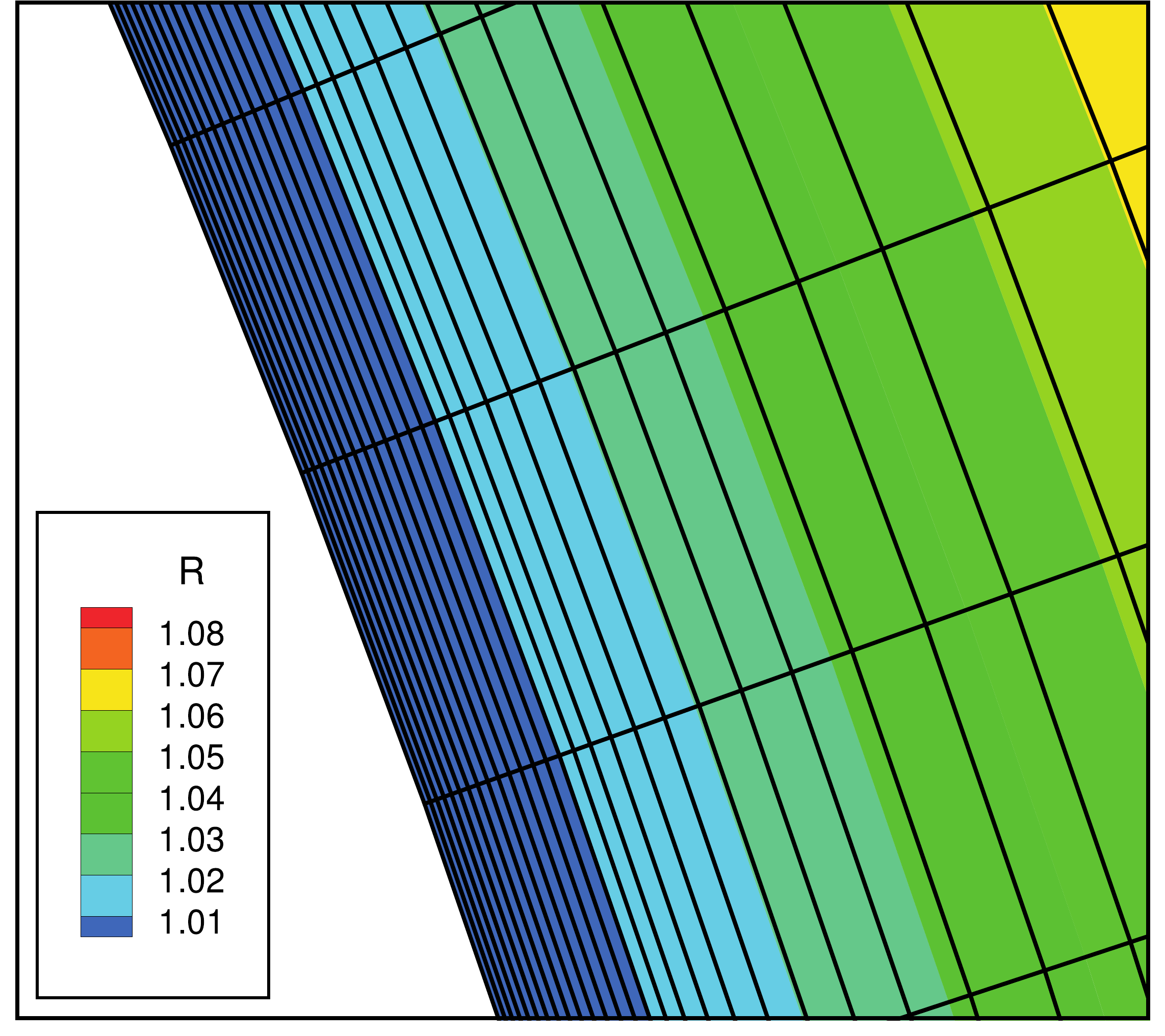}
\hspace{0.5in}
\caption{\small Comparison of grid geometry and cell sizes near the solar surface for the SWMF SC model (coloring indicates radial extent). Left: a Cartesian grid with minimum $ds = 6,800\ \text{km}$. Right: The customized spherical grid with minimum $dr = 230\ \text{km}$ (part of this work). All three directions do not have to be further refined at the surface to enhance the radial resolution when using spherical geometry, leading to significant savings in computational overhead. In this example, both simulation domains have comparable numbers of total cells.}
\label{fig:geometry} 
\end{figure}

To address this scale and resolution issue we adapted the generalized grid capability of the SWMF, which has ability to calculate MHD fluxes in arbitrary geometries, to construct a spherical (r$, \theta, \phi$) grid function with highly non-uniform radial scales near the transition region ($dr = 230$ km) that smoothly transitions to near equal face area at large $r$ ($dr = 30,000$ km at $r = 5 R_\sun$). Because the grid function is continuous, the model maintains the block-adaptive mesh and adaptive mesh refinement (AMR) capability for flexible local spatial resolution both initially and as the simulation progresses. For this study, focusing on the global structure of the corona, we use a uniform spacing in the angular directions with $d\theta = d\phi = 1.4^\circ$ at the surface, which is coarsened by a factor of two via AMR beyond $1.7 R_\sun$ for regions within $\pm65^\circ$ and $1.2 R_\sun$ outside (to avoid needlessly small cell sizes at the poles). The outer boundary is fixed at $r=24\text{R}_\sun$. Degenerate cells touching the polar axes are treated with azimuthal averaging to avoid discontinuity across the poles. A grid comparison near the surface between this work and an AMR cartesian mesh with a similar number of total cells is shown in Figure \ref{fig:geometry}.

\subsection{LOS image synthesis}To study the EUV corona and associated dynamics in the context of real events, we include a proper treatment of integrated coronal emission in the EUV range, particularly for the purpose of comparing simulated observations to existing EUV observations by a current solar observatory. This process, following the work of \citet{mok05,mok08,lionello09}, involves two main steps: (i) characterization of detector response based on physical parameters that are calculated in the model and (ii) line-of-sight (LOS) integration though the 3D data set to create synthesized image.
\\
\label{section:LOS-synthesis}

The first step is to understand and calculate the instrumental response for coronal material at a given temperature and density. For simplicity and ease to compare to studied observations we chose to model the three coronal band-passes at 171, 195, and 284\AA \ of the EIT instrument on the SOHO spacecraft \citep{delaboudiniere95} and the AlMg filter of the SXT instrument \citep{tsuneta91}.  Because of the extremely high temperature and low densities, the emission line EUV and Soft X-Ray radiation of the solar corona above ~0.5 MK can be treated as optically thin to a good approximation. As such, the intensity measured by each pixel of an imaging detector can be treated as a LOS integral through the coronal plasma, namely:
\begin{equation}
R = \int{n_e^2 f_i(T,n_e)dl}  \text{ DN } \text{s}^{-1}
\label{los}
\end{equation}
where $R$ is the pixel response measured in Data Numbers (DN) per second. Here $l$ follows along the line-of-sight, the square of the electron density, $n^2_e$, accounts for the amount of emitting material and $f_i(T,n_e)$ is the instrumental response of filter $i$ per unit emission measure as a function of both temperature and density. This is equivalent to integrating the resolved Differential Emission Measure (DEM) distribution along $l$. To calculate the response functions, $f_i(T,n_e)$, we use the CHIANTI version 5 emission line analysis routines \citep{landi06} to generate synthetic spectra, and the EIT and SXT analysis routines for detector response (a part of the SolarSoft (SSW) framework written in IDL \citep{freeland98}).

\section{Model Runs}
\label{section:Model_Runs}

In this section we present the application of the LC model to CR 1913, initialized using a synoptic MDI magnetogram centered on Aug 27, 1996. Occuring during solar minimum, the sun at this time  features a conspicuous equatorial coronal hole extending from the north to down past disk-center, as well a large active region. This then conveniently allows us to examine the ability of a given model to describe these three basic regimes: the average quiet-sun, coronal holes, and active regions. Additionally, this rotation has been studied in detail using similar methods (e.g. \citet{mikic99,lionello09}) and thus provides a means of comparison to existing work.
\\

\begin{figure}[hbtp]
\centering

\includegraphics[width=0.80\textwidth]{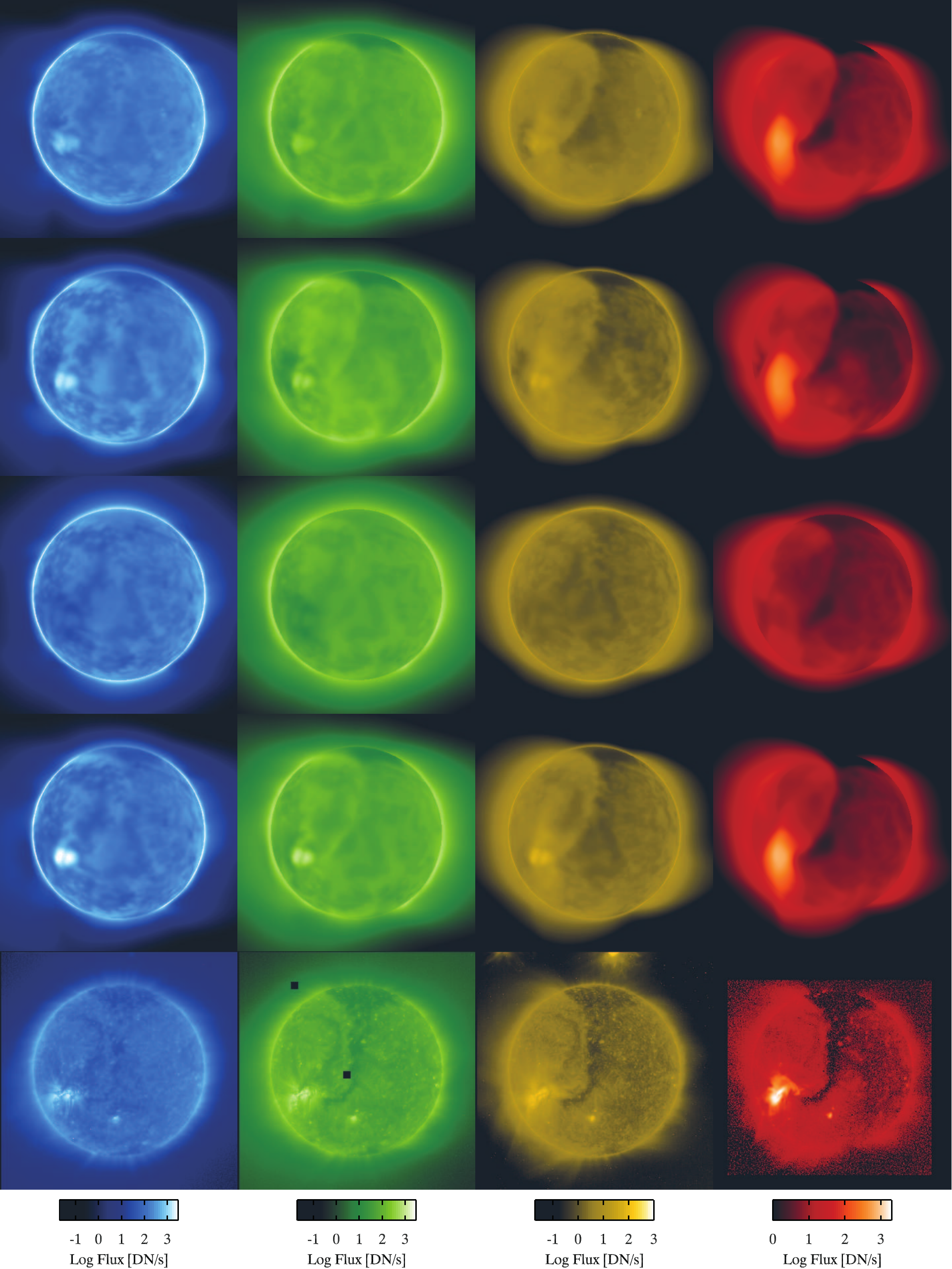}
\caption{\small Comparison of EIT 171\AA, 195\AA, 284\AA, and SXT AlMg image synthesis to observations for the model runs of CR 1913 centered on Aug 27, 1996. Runs A, B, C, and D are shown top to bottom. Bottom row: SOHO EIT and SXT Almg observations of the same date near 01:00 UT (00:10:13, 00:24:05, 01:05:19, and 01:07:28 respectively). Run A: REB BC + B weighted heating + Open Field Modification. Run B same as A with chromospheric BC. Run C: Chromospheric BC + uniform exponential heating. Run D: chromospheric BC + Exponential heating + B weighted AR component + open field modification.}
\label{fig:EUVcomp_4runs} 
\end{figure}

Four model runs used to demonstrate the usefulness of the LC model are summarized in Table \ref{table:model_runs}. As a convenient comparison of their global thermodynamic and topological properties, the LOS synthesis of EUV and soft X-Ray emission are shown in Figure \ref{fig:EUVcomp_4runs}. Each of the models presented use either $|B|$ weighted heating (model 1, section \ref{heating_model_1}) or exponential heating (model 2, section \ref{heating_model_2})). All four use coronal hole heating  (section \ref{heating_model_CH}) applied uniformly to the domain, while all but run C use the open/closed field weighting (section \ref{heating_model_OC}) to achieve a more realistic equilibrium in these regions. In order to include strong heating in Active Regions (AR) with high magnetic field strength, Run D also includes a modification to the heating function, and transitions smoothly from exponential heating to $|B|$ weighted heating ($Q_h = 4\times 10^{-5} |B|\ \text{ergs cm}^{-3}\text{s}^{-1}$) above $30$ Gauss.
\\

\placetable{table:model_runs}

\begin{deluxetable}{cccccc} 
\tablewidth{0pt} 
\tablecaption{\small Model Runs.\label{table:model_runs}} 
\tablehead{\colhead{Run} & \colhead{BC} & \colhead{Heating} & \colhead{Open Cutoff} &
\colhead{Power [ergs $\text{s}^{-1}$]}}
\startdata 
 		A & REB & $\propto |B|$ & Yes & $2.88\times\text{10}^{28}$\\ 
		B & Chromo & $\propto |B|$ & Yes & $2.88\times\text{10}^{28}$\\
 		C & Chromo & Exp & No & $2.49\times\text{10}^{28}$\\
 		D & Chromo & Exp + AR & Yes & $2.88\times\text{10}^{28}$\\
\enddata
\end{deluxetable}

All four model runs show similar global features in the EUV and Soft X-Ray when compared to observations but differ on the precise details. For example, all four runs resolve cool temperatures along the the polarity inversion lines near the surface (seen to the east and west of the extended equatorial coronal hole near the center of the disk). This is naturally due to field line orientation and loop length governing thermodynamic equilibrium. This commonality serves to emphasize the strong role that the 3D magnetic field topology of a given Carrington Rotation plays in determining dynamic equilibrium in the corona.
\\

In the comparison between Run A (REB model BC) and Run B (Chromospheric BC) we observe more uniform emission near the surface as well as slightly higher temperature (as seen in the longer extent of SXT emission) for the REB case. Because the boundary is near the top of the transition region in the REB model, the height of the transition region is less able to vary in response to magnitude of the downward heat conduction flux from above. Also, since the REB model does not include a region with chromospheric densities (where the radiative cooling is much higher due to the $n_e^2$ dependence) a smaller fraction of the coronal heating power is lost in the transition region, leading to higher equilibrium temperatures. However, it is clear that the REB model, which is more suited for computational efficiency due to the longer equilibrium length scales at $T= 500,000\text{K}$, can be used to adequately describe the ambient global corona.
\\

One of the most striking results is perhaps the comparison of Run C to the other model runs. In this run, the coronal heating function was applied uniformly in the computational domain (no dependence on $\theta$ or $\phi$). While the coronal holes and significant AR emission are unsurprisingly not well preserved in the EUV, still much of the basic features of the low corona are reproduced. The ability of simple uniform heating model to reproduce obvious features on the surface, such as average quiet sun emission, and lessening near inversion lines, as well as the bi-modal structure between open and closed field regions at higher temperatures ($284\text{\AA}$ line and soft X-Rays) speaks to the importance of both the 3D magnetic topology in determining the dynamic equilibrium of the corona and the redistribution of energy via a thermodynamic energy equation including electron heat conduction.
\\

\begin{figure}[htbp] 
  \centering
   \includegraphics[width=0.42\textwidth]{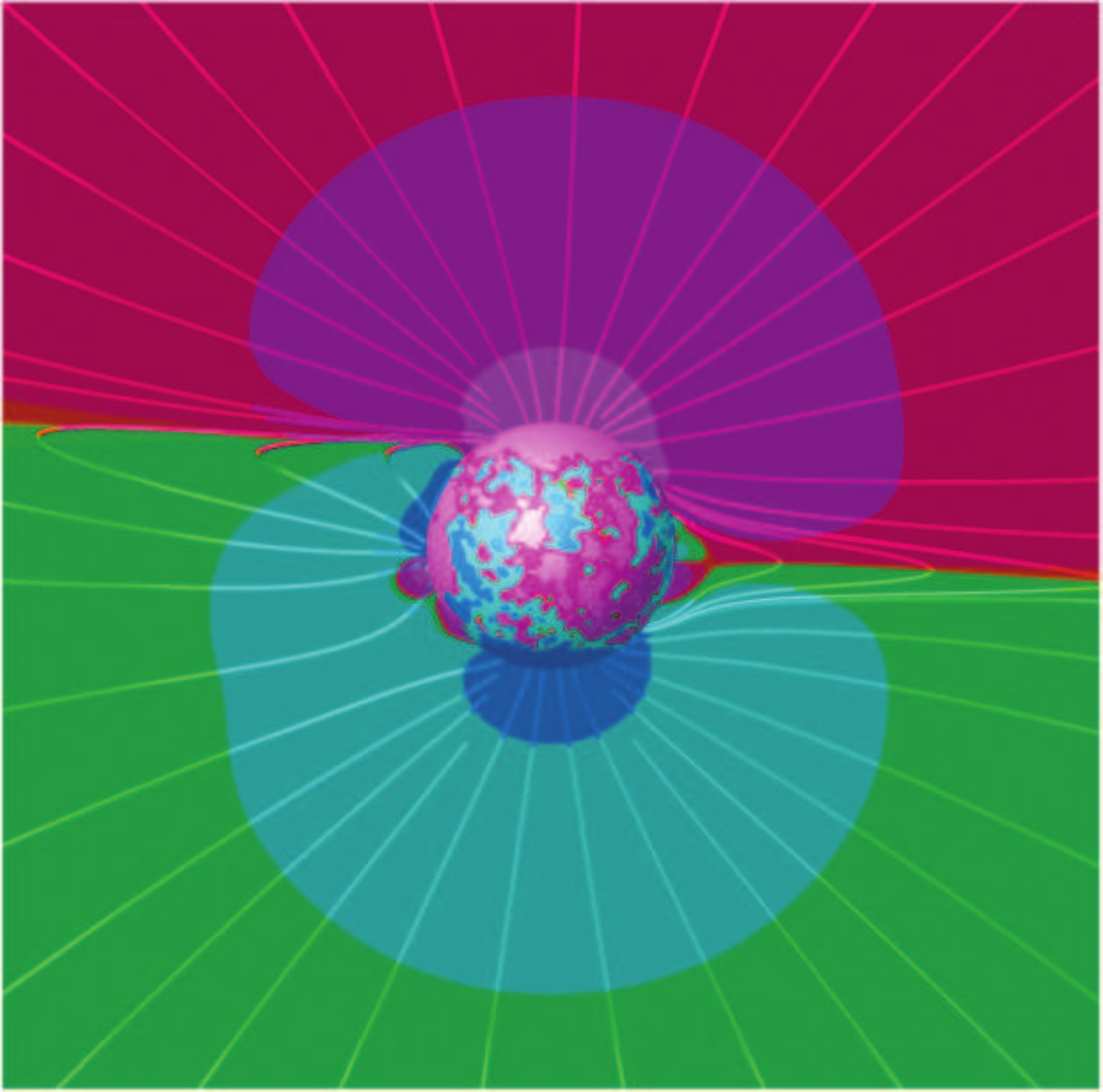}  
   \includegraphics[width=0.42\textwidth]{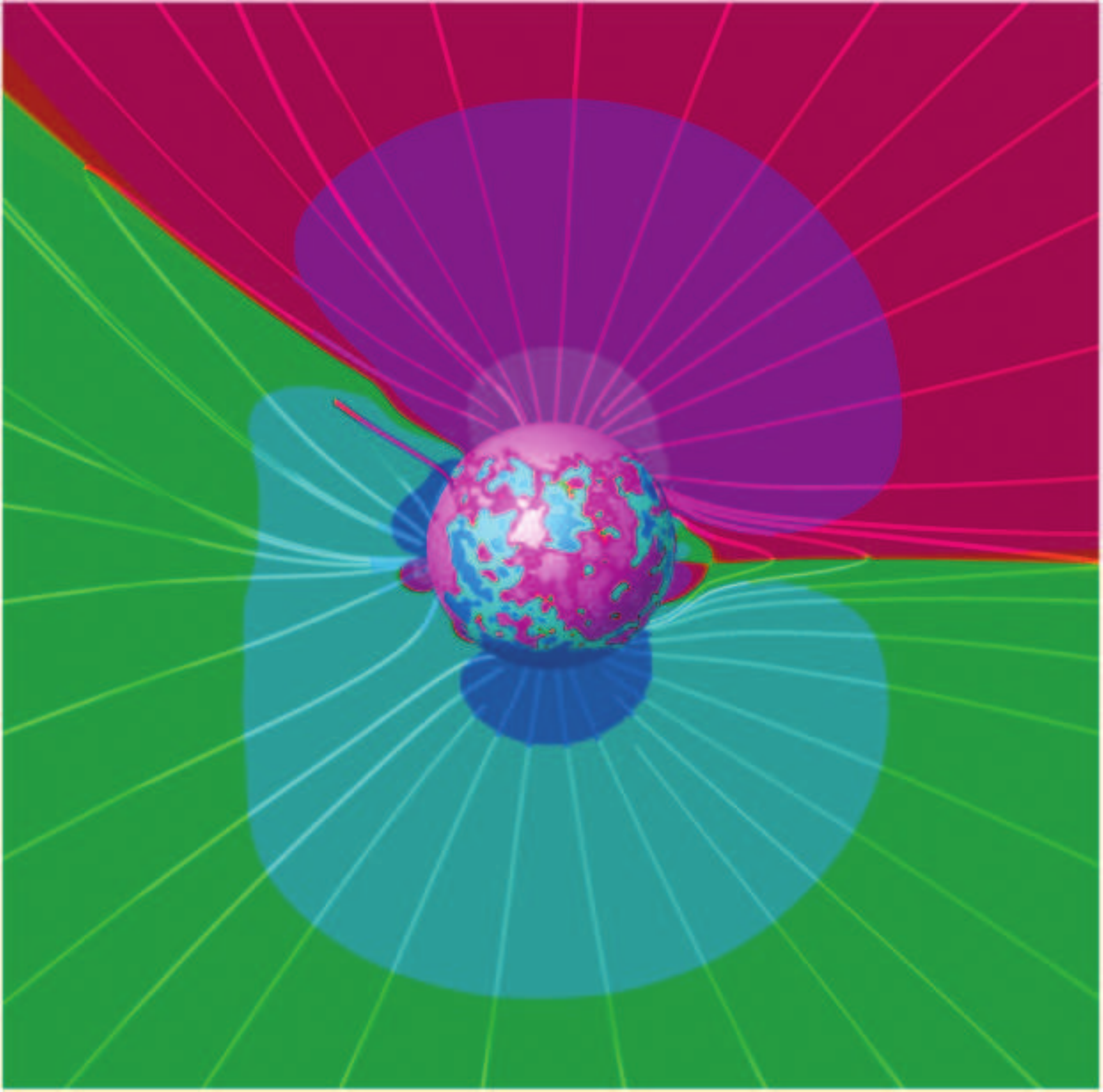}
   \includegraphics[width=0.08\textwidth]{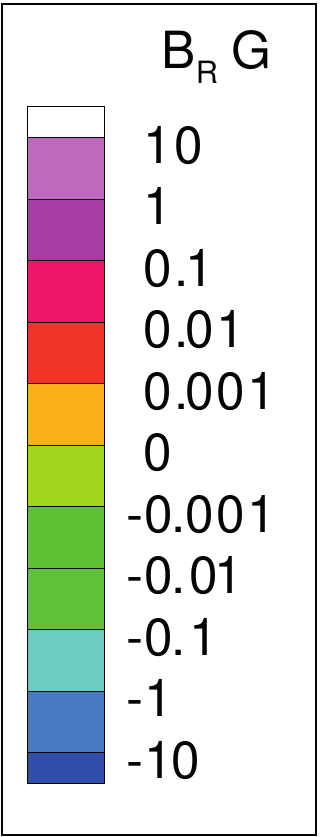}
   \caption{\small Comparison of the significant effect that AR heating has on the tilt of the heliospheric current sheet between Run C (uniform heating, left) and Run D (with AR heating and open field modification, right). The thermodynamic stress induced by significant local heating of the active region ($\sim 7\times10^{27} \text{ergs s}^{-1}$) causes an extreme shift in the northward tilt of the heliospheric current sheet $\sim 35^\circ$. Simultaneously, the region not thermally connected to the active region on the opposite side of the sun remains unchanged. The color contours display the radial component of the magnetic field, $B_r$. The chosen slice is parallel to the polar axis and intersects the inversion line between the large active region seen on Aug 27, 1996. The 3D magnetic streamlines intersect the slice at the same location in each figure.}
   \label{fig:streamer_Br_slice}
\end{figure}

For a more realistic empirical model in Run D, we modify Run C by including AR heating and the open-field cutoff to better describe heating in active regions in coronal holes. The most interesting finding in this comparison is that by including significantly enhanced heating in the large active region, we observe a pronounced effect on the equilibrium structure of the AR associated closed field/streamer region (Eastern side of the corona in Figure \ref{fig:EUVcomp_4runs}). This produces noticeable feedback on the global structure as seen via a larger northward tilt of the heliospheric current sheet near the AR longitudes (shown in detail Figure \ref{fig:streamer_Br_slice}), and suggests a complex relationship both between the thermodynamics of the low corona and the global structure of the solar wind. Because this effect cannot be fundamentally extracted from extrapolating the magnetic field alone, this provides not only a stong motivation for using thermodynamic models, but also another element with which to constrain and refine AR heating models in future studies (the one used in in this work being a necessary but crude approximation).
\\

Another important inference about coronal heating can be gleaned from the comparison of the two heating models directly (Run B and Run D). Although the total heating power is the same in each model, the $|B|$ weighted heating model naturally has a shorter scale height (due to the falloff of $|B|$ with radius) and we subsequently observe slightly enhanced pressure in the low corona and lower temperatures (hence density scale height) in the closed field regions. However, the fact that the simple exponential heating model applied to the quiet sun in Run D can reproduce the mean structure of the low corona with the same fidelity as a more complicated empirical model, demonstrates that perhaps complex empirical models are not as entirely necessary as some might suggest and motivates a need for more physics based models to advance further.

\section{Detailed Analysis}
\label{section:Detailed_Analysis}

\label{topological analysis}

In this section we provide a more detailed analysis of Run D in particular. We examine the magnetic and thermodynamic structures realized by the LC model and provide a comparison to the standard SC model \citep{cohen07}.

\subsection{Temperature structure of low corona} 
\begin{figure}[htbp] 
  \centering

   \includegraphics[width=0.8\textwidth]{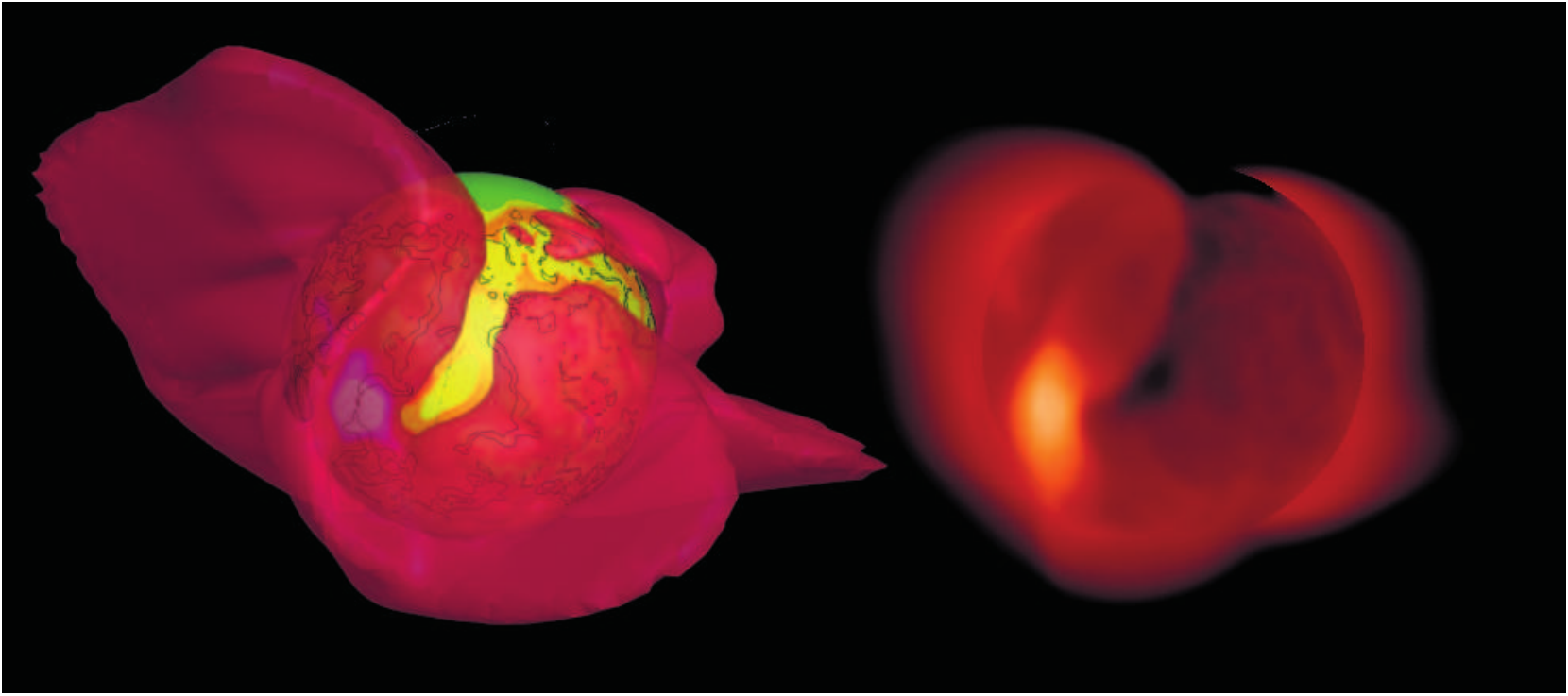}
   \includegraphics[width=0.13\textwidth]{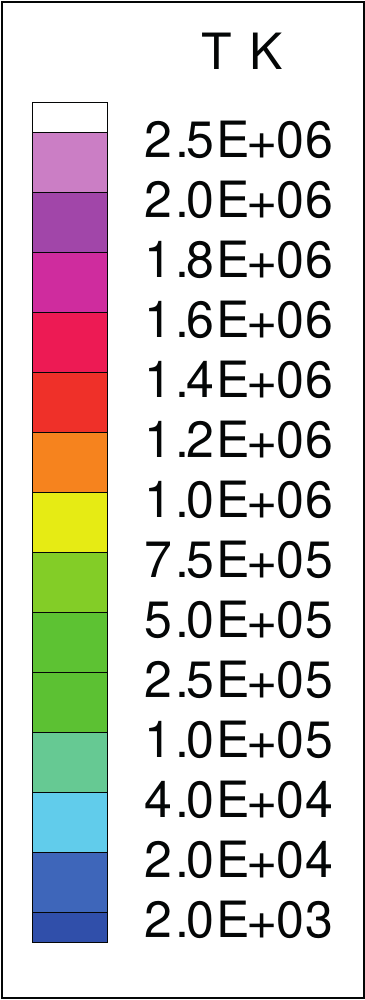} 

   \caption{\small Visual comparison of 3D topological structure (left) to LOS synthesis of SXT AlMg response (right) for Run D (Exponential heating + AR heating).The surfaces shown are of constant $r=1.03\ \text{R}_\sun$ and $T=1.6\ \text{MK}$ with color contours of T. The black contour lines of normal radial field direction $\hat{|B_r|}=0.2$ are shown to illustrate how the hot, closed field streamer regions overlay the inversion lines on the surface.}
   \label{fig:Temperature_topology}
\end{figure}

An important advantage of including heat conduction in a global 3D enrionment is the ability to the study complex open/closed field topologies in a self consistent manner, which allows one to study the feedback effect that the magnetic topology has on heat distribution and vice versa. The most striking effect can be seen by examining the differences between heat conduction along open or closed field geometries.  Because closed field regions represent closed systems for the flow of thermal energy, they acheive higher temperatures than their open-field counterparts and shift into the soft-Xray regime ($T\sim>1.5$MK). This natural correspondence can be clearly seen in Figure \ref{fig:Temperature_topology} where we show a 3D surface at fixed temperature ($T_e = 1.6$ MK) and an LOS image of soft-X-Ray emission. 

\subsection{Magnetic structure of low corona} 
\begin{figure}[htbp] 
  \flushleft
   \includegraphics[width=0.30\textwidth]{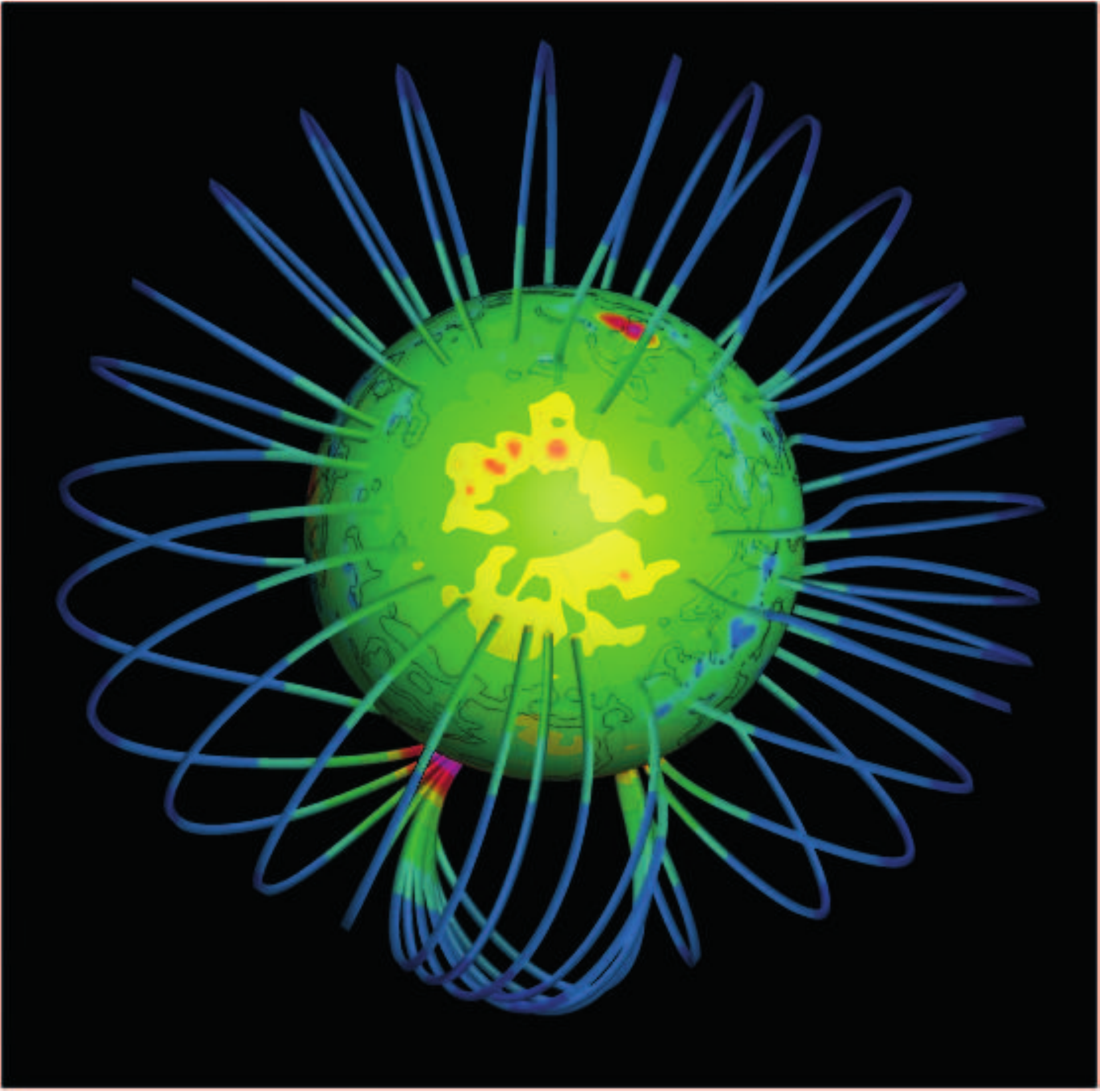} 
   \includegraphics[width=0.30\textwidth]{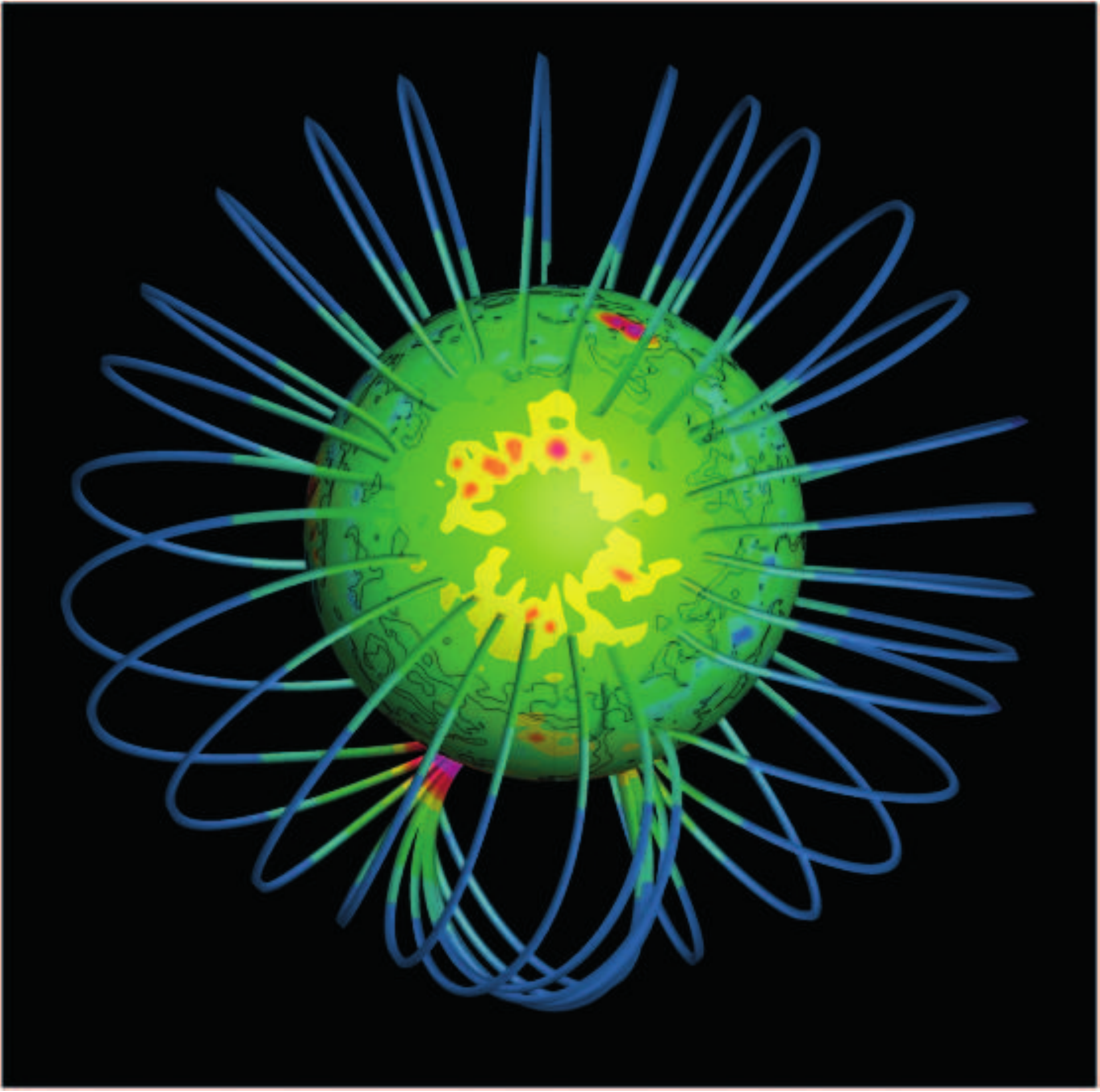} 
   \includegraphics[width=0.30\textwidth]{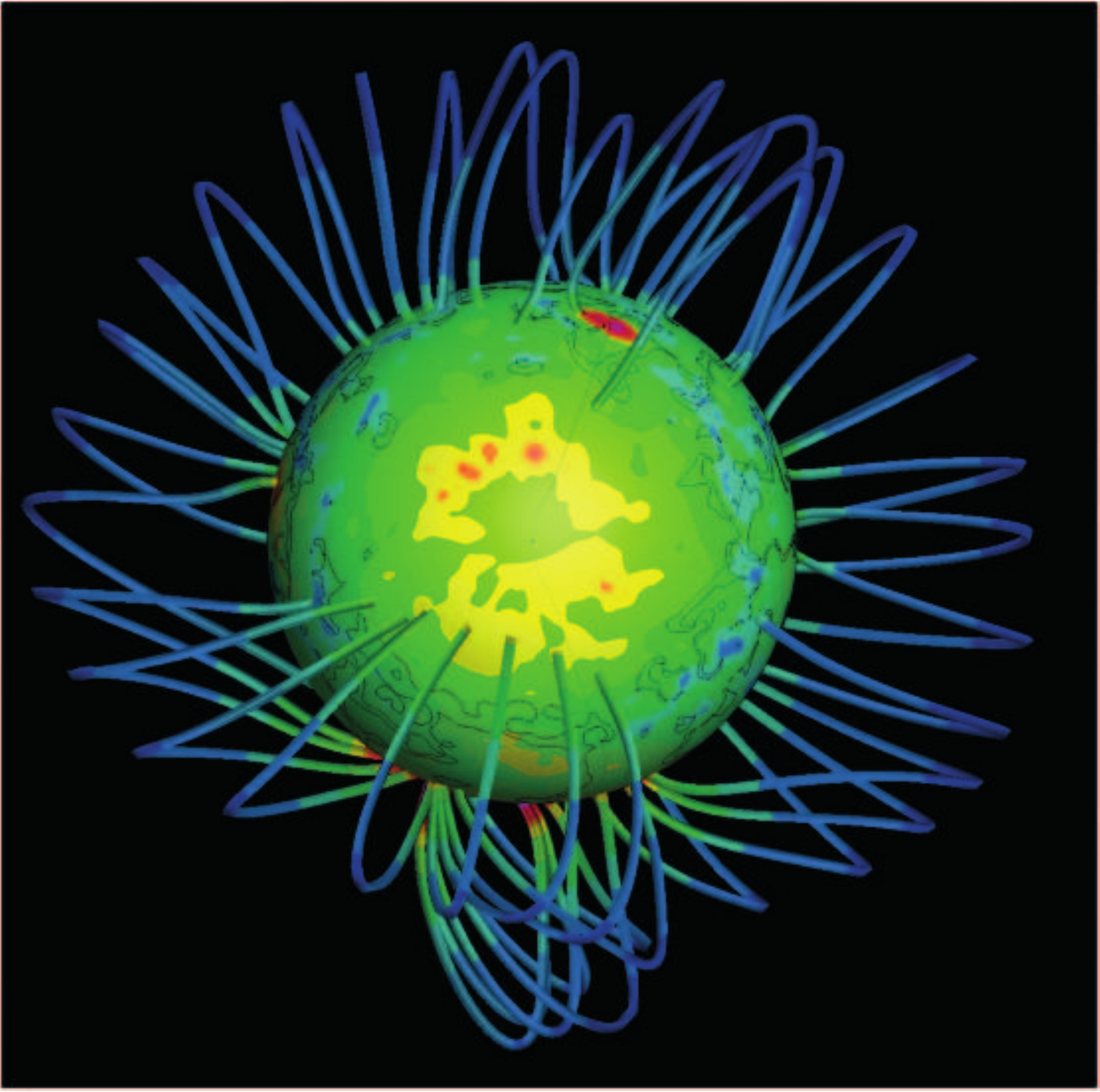} 
   \includegraphics[width=0.30\textwidth]{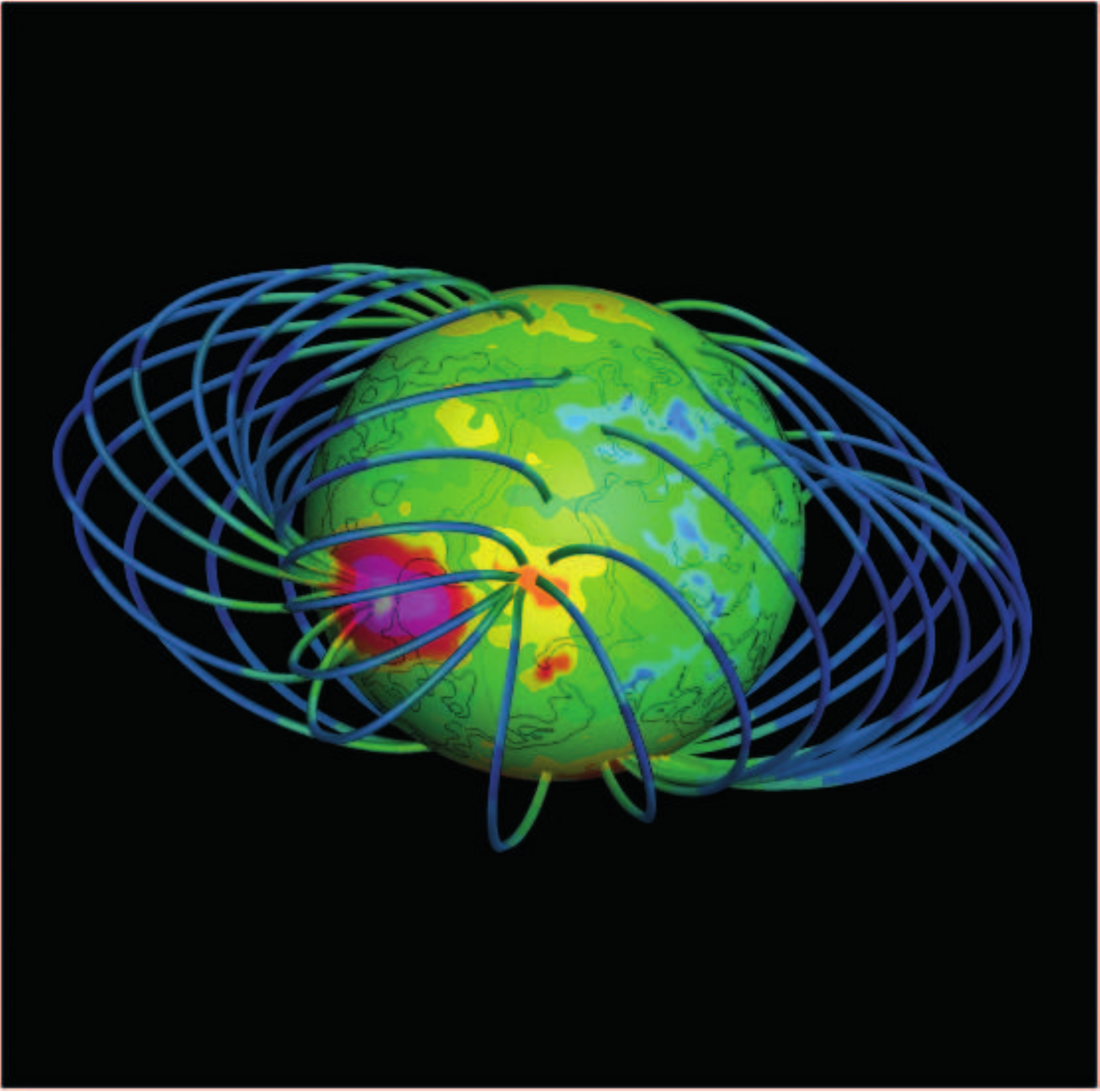} 
   \includegraphics[width=0.30\textwidth]{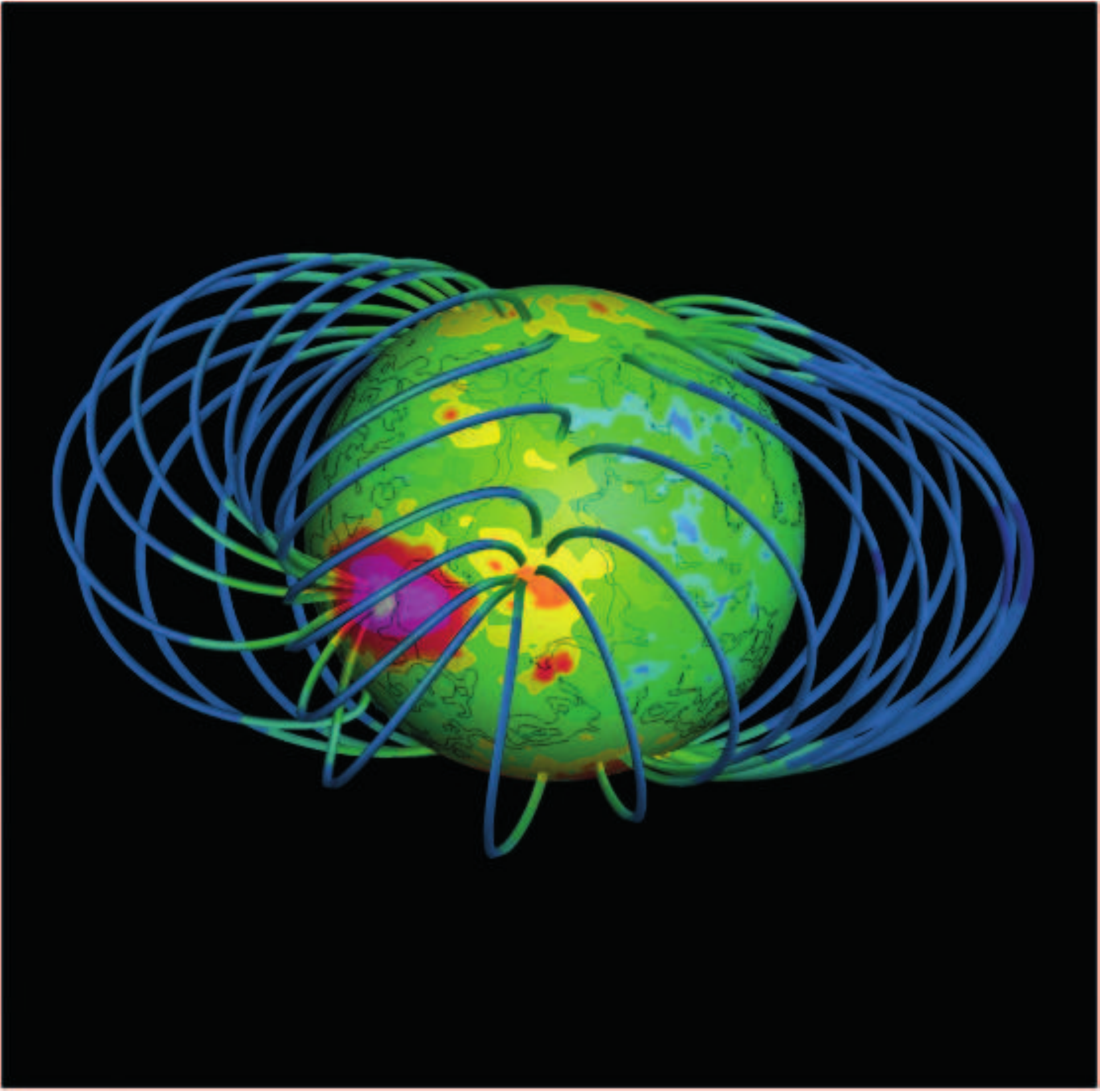} 
   \includegraphics[width=0.30\textwidth]{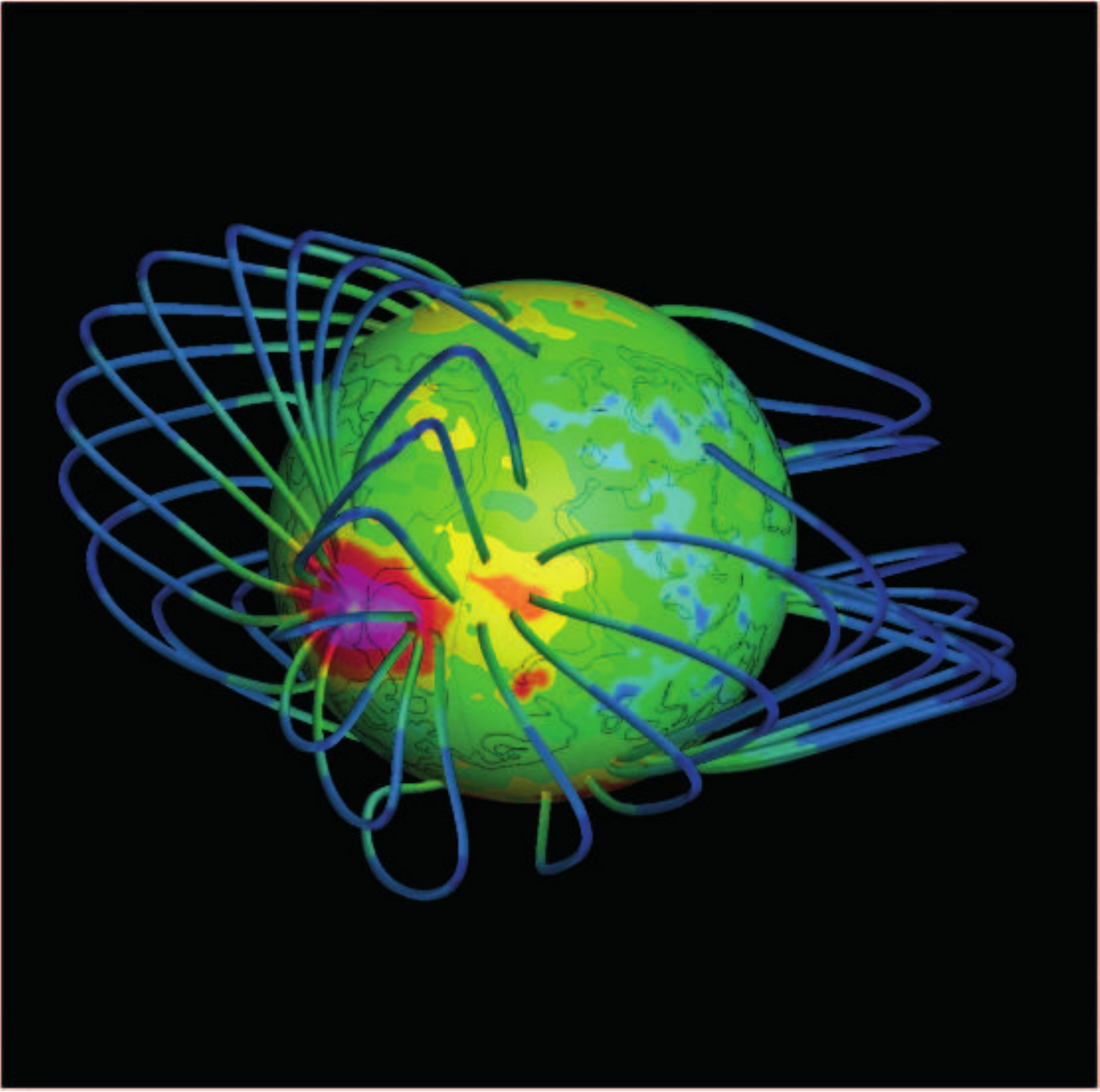}
   \includegraphics[width=0.07\textwidth]{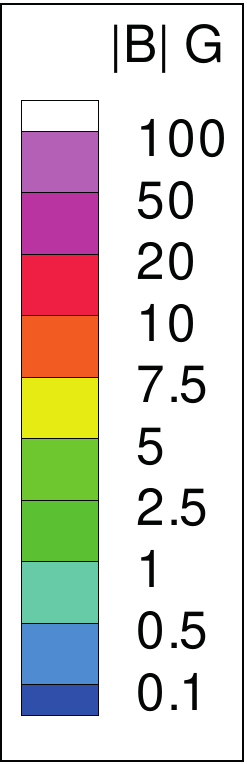} 
   \caption{\small Magnetic field comparison of the streamer structure of CR 1913 between 3 different models, from a top down (top) and Earth-centered viewpoint (bottom) at 01:00 UT Aug 27 1996. Left: The PFSSM initial condition \citep{altschuler77}. Middle: Steady state equilibrium of the default SC model \citep{cohen07}. Right: Steady state of Run D (Exponential heating + AR heating). It is immediately clear that the closed field structures become highly stressed by heating near the surface and the subsequent redistribution via heat conduction. Selected field lines are chosen at near equal intervals along the intersection of $r=2.0 \text{R}_\sun$ and $\hat{|B_r|}=0.5$ in each model.}
   \label{fig:Magnetic_topology}
\end{figure}

It is also important to examine in detail the effect that the thermodynamic model has on the magnetic structure as a whole. In Figure \ref{fig:Magnetic_topology} we show a comparison between closed field lines at $r=2.0\text{R}_\sun$ for the PFSSM initial condition \citep{altschuler77}, the standard SC model \citep{cohen07} and Run D. Immediately obvious are the drastic changes between the closed field/streamer structure of the polytropic SC model and the thermodynamic LC model. Unsurprisingly, the high temperatures and thermal energy introduced by coronal heating and conduction along the closed field produces a large thermal stress in the relatively high $\beta$ regions near the streamer cusps along the current sheet. This in turn leads to significant reorganization of the magnetic field as compared to a PFSSM or polytropic MHD model. The fact that this sort of stress is highly sensitive to both the local nature of a given coronal heating model (Figure \ref{fig:streamer_Br_slice}) and the global magnetic topology (Figure \ref{fig:Magnetic_topology}) is a leading motivation for using a fully 3D, thermodynamic model when studying the structure of the global corona.

\subsection{EUV and Soft X-Ray comparison}
\begin{figure}[hbtp]
\centering
\includegraphics[width=0.90\textwidth]{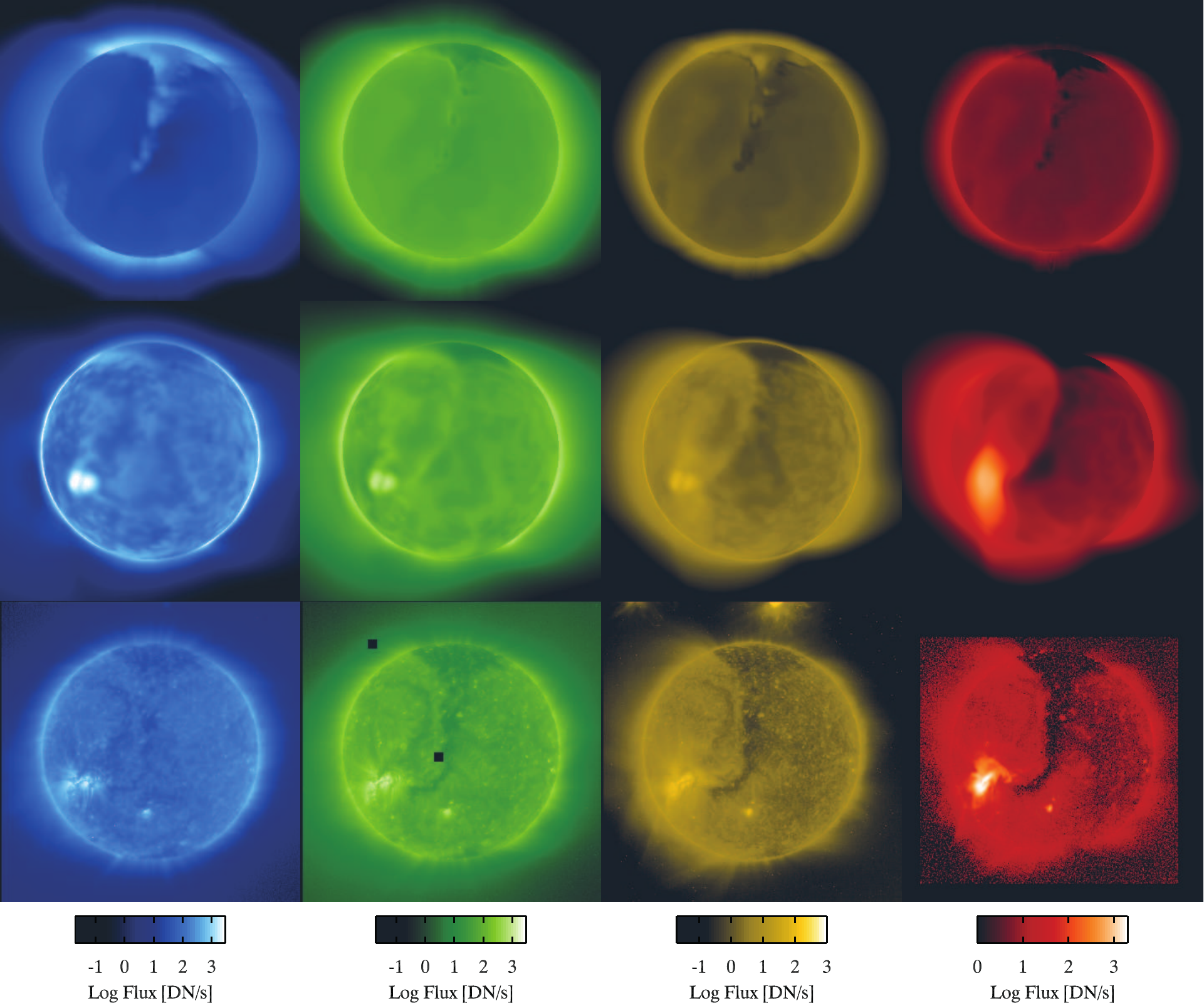}
\caption{\small Comparison of EIT 171\AA, 195\AA, 284\AA, and SXT AlMg image synthesis to observations for Aug 27, 1996. Top row: steady state of the standard SC model \citep{cohen07}. Middle row: Run D (Exponential heating + AR heating), demonstrating significantly improved agreement with surface conditions and streamer topology. Bottom row: SOHO EIT and SXT Almg observations of the same date near 01:00 UT (00:10:13, 00:24:05, 01:05:19, and 01:07:28 respectively).}
\label{fig:EUVcomp_run10toWSA} 
\end{figure}

\begin{figure}[hbtp]
\centering
\includegraphics[width=0.30\textwidth]{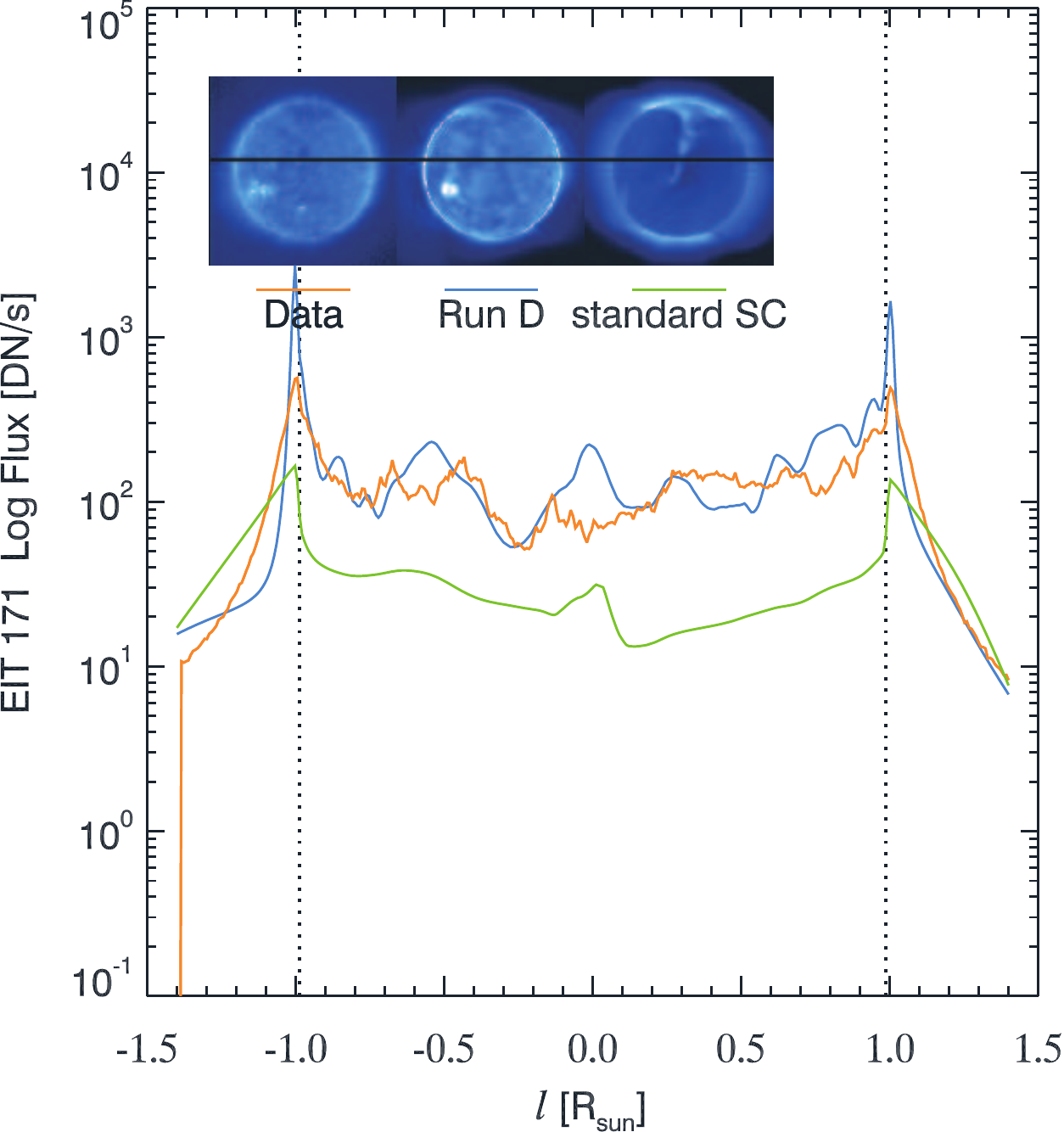}
\hspace{0.8in}
\includegraphics[width=0.30\textwidth]{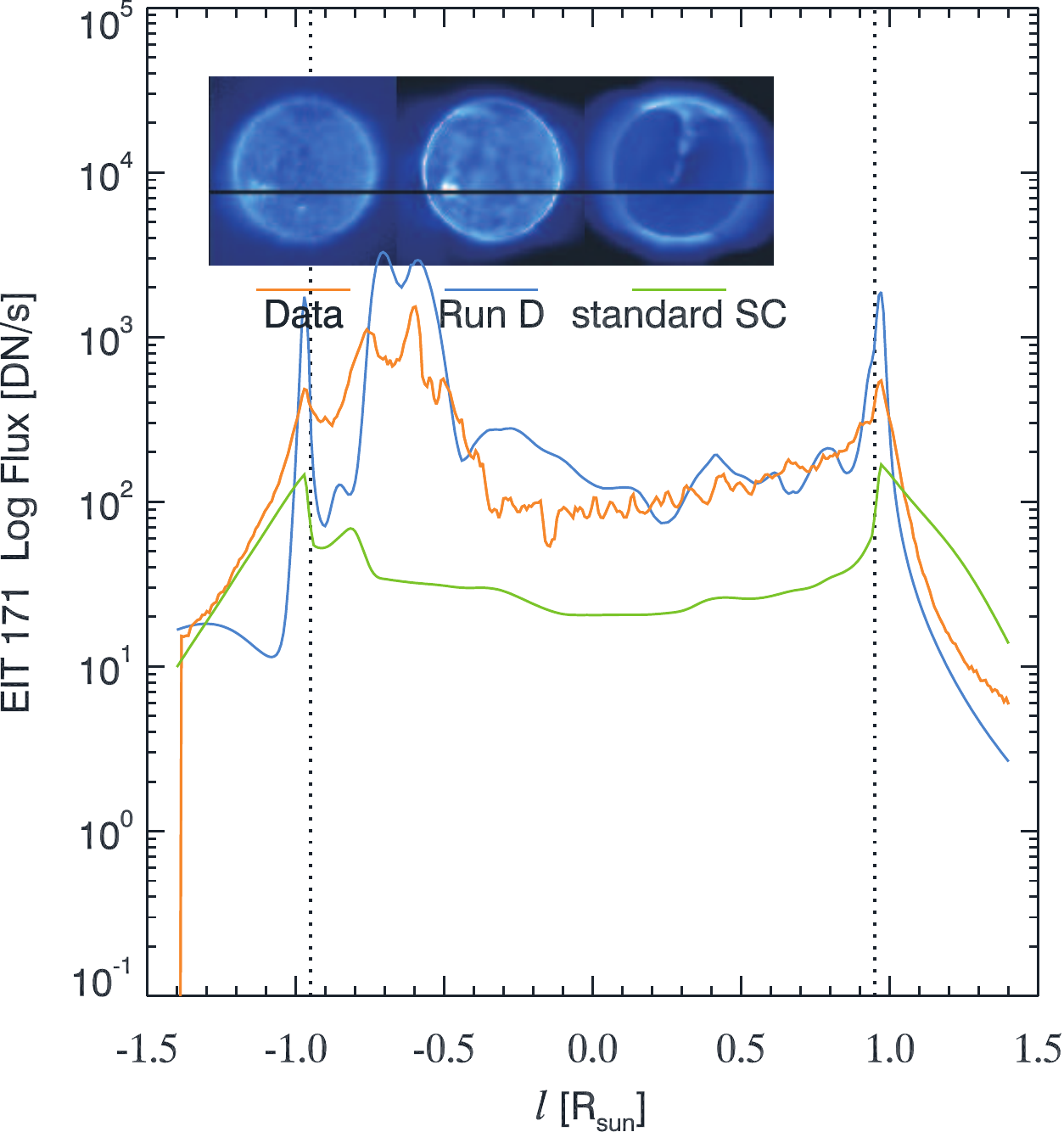}
\includegraphics[width=0.30\textwidth]{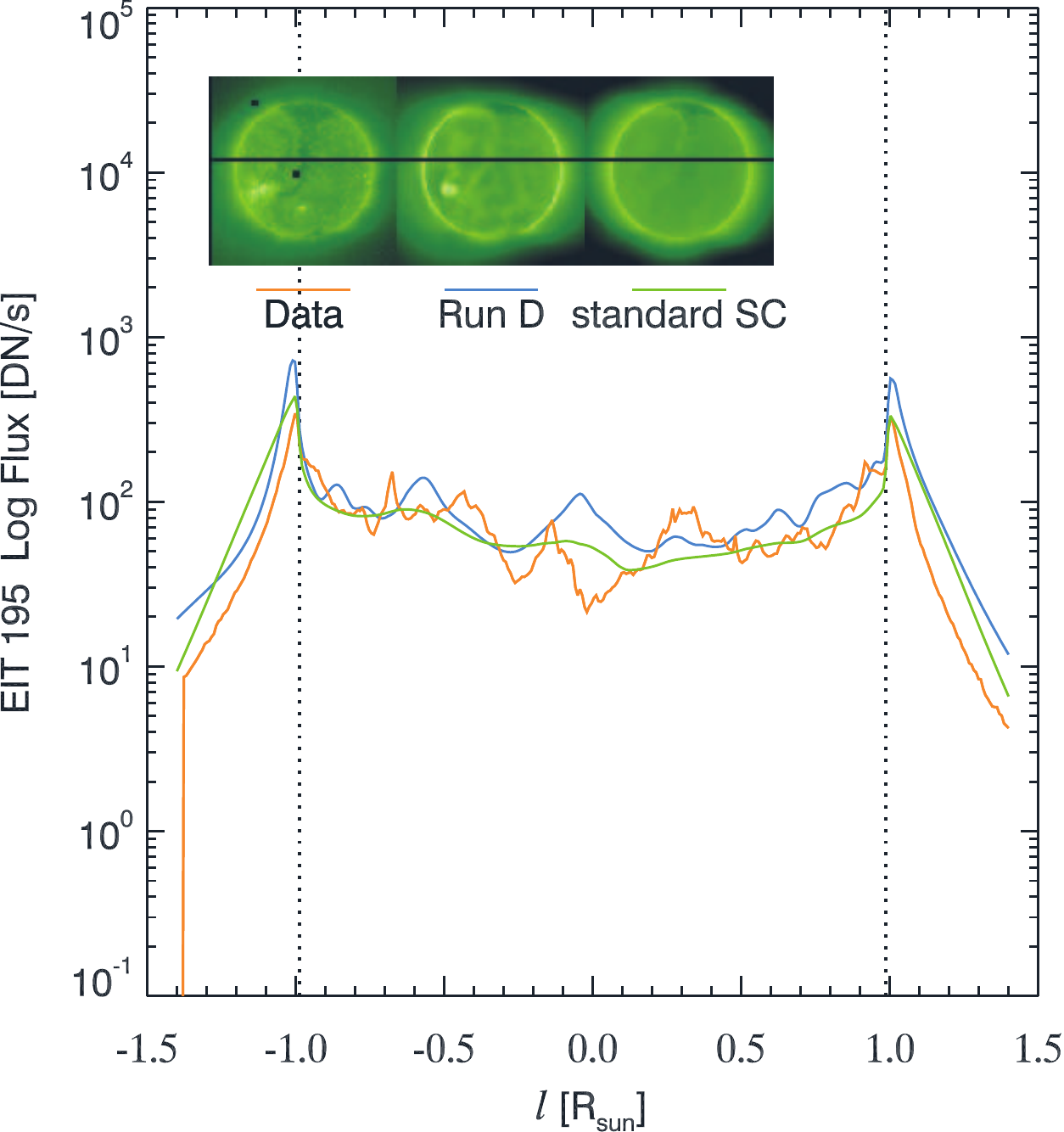}
\hspace{0.8in}
\includegraphics[width=0.30\textwidth]{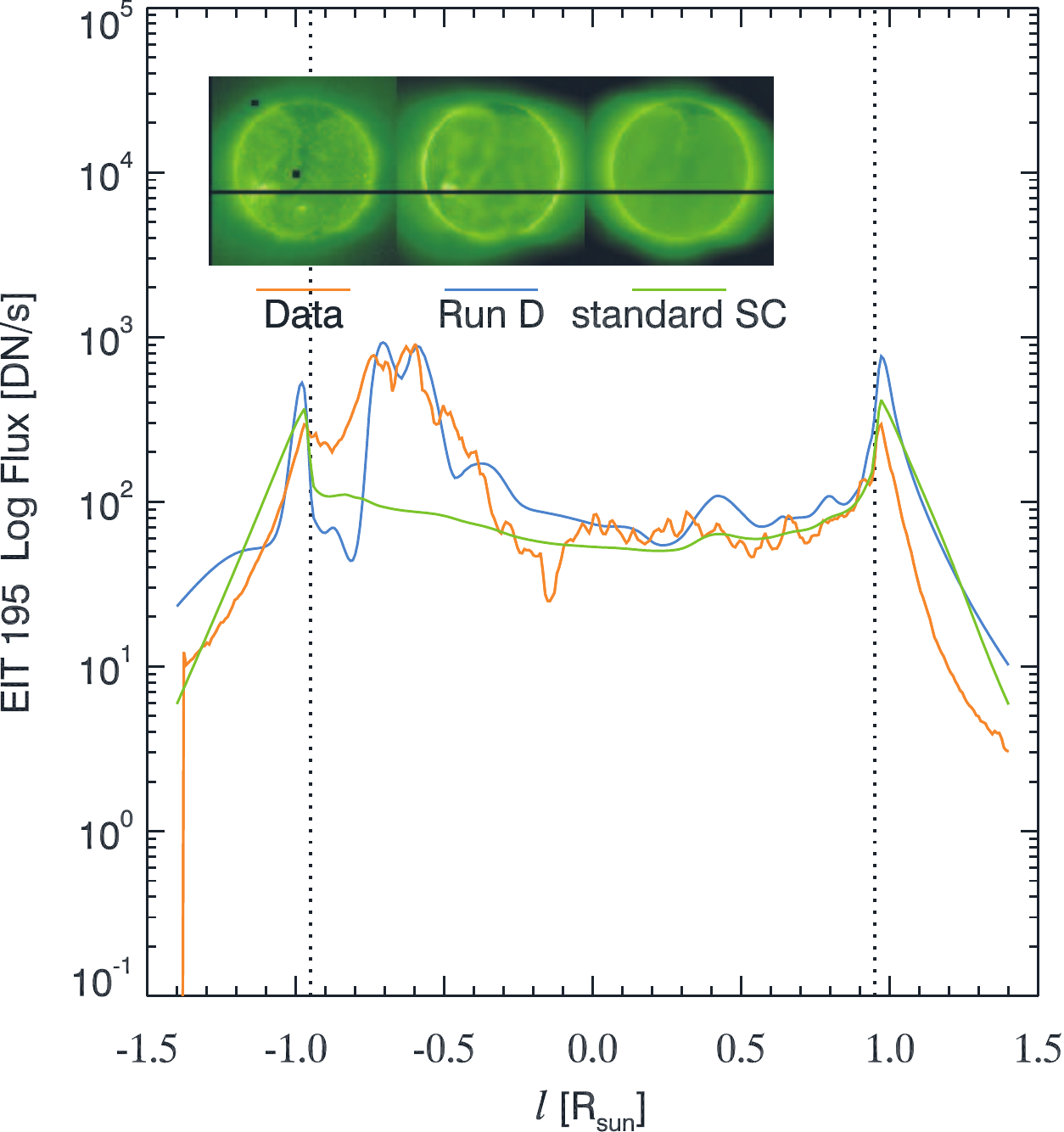}
\includegraphics[width=0.30\textwidth]{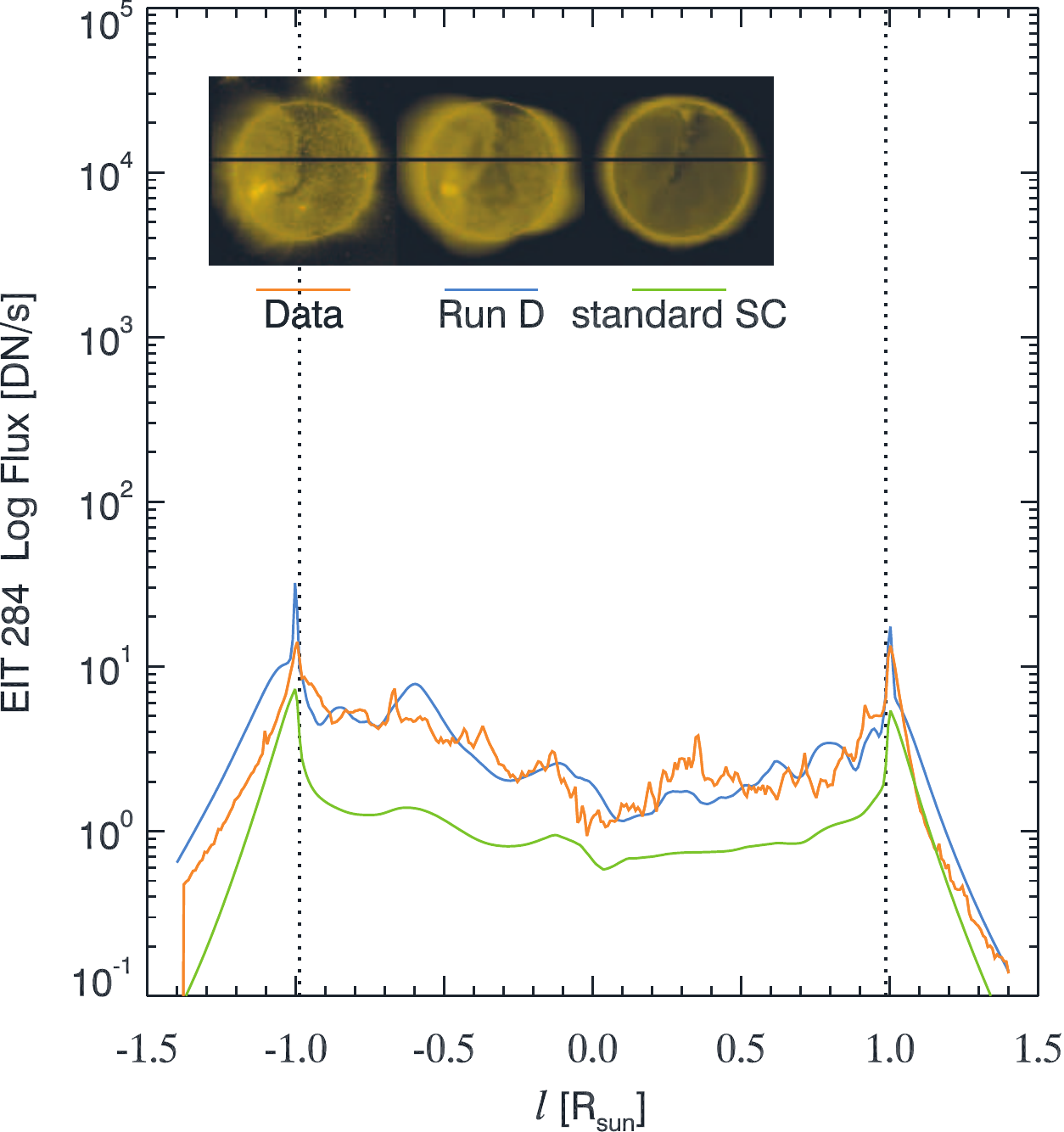}
\hspace{0.8in}
\includegraphics[width=0.30\textwidth]{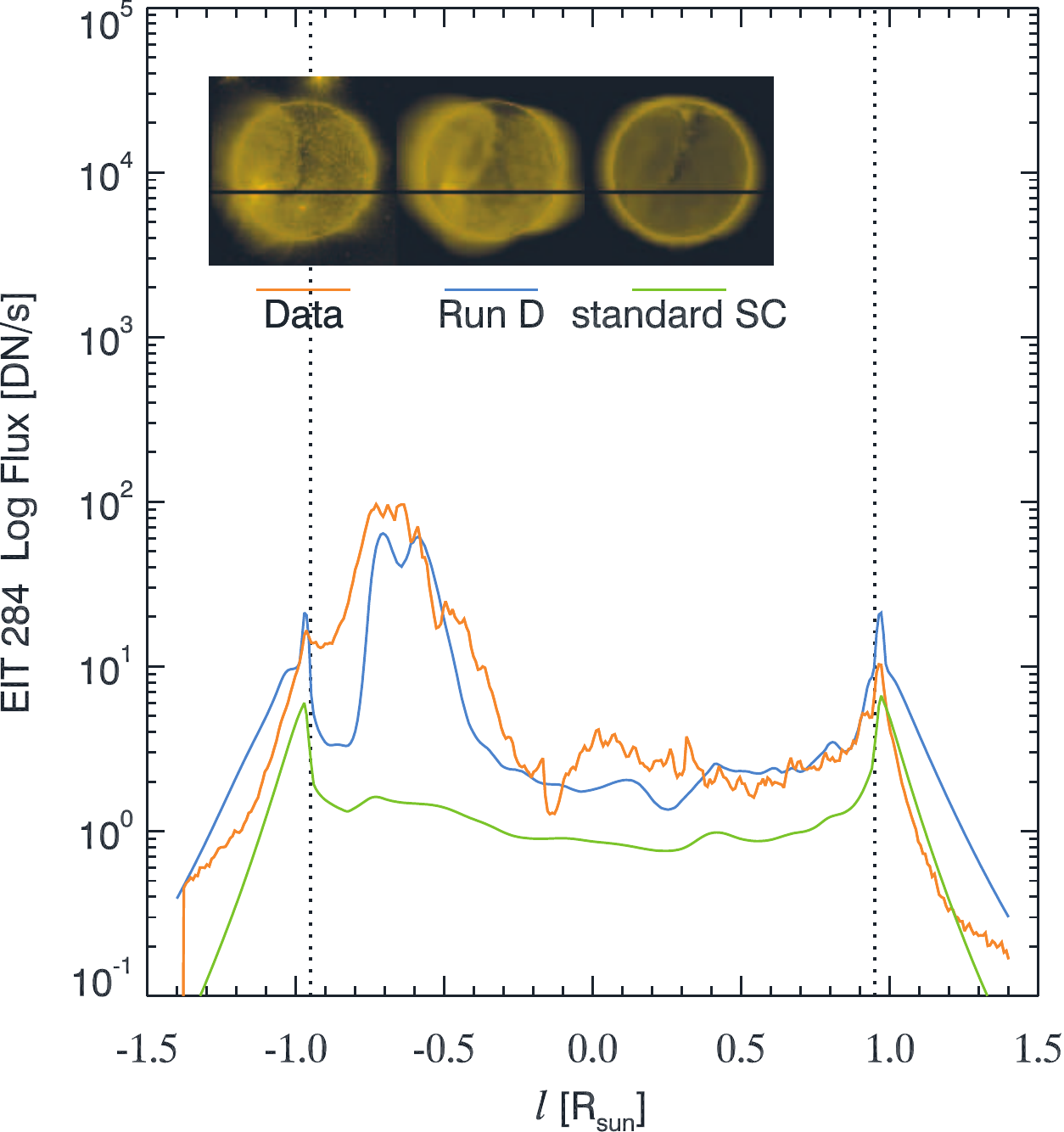}
\includegraphics[width=0.30\textwidth]{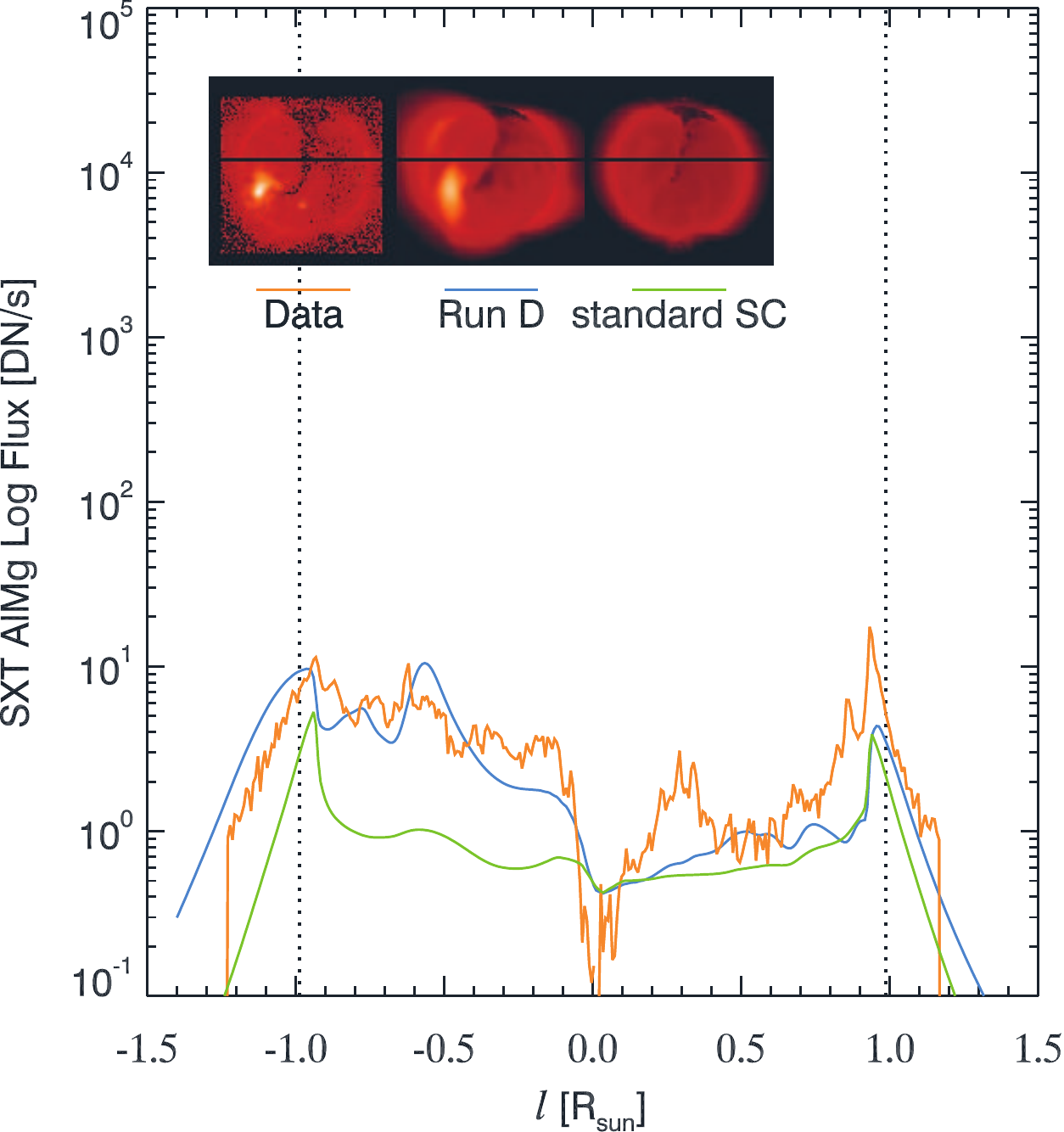}
\hspace{0.8in}
\includegraphics[width=0.30\textwidth]{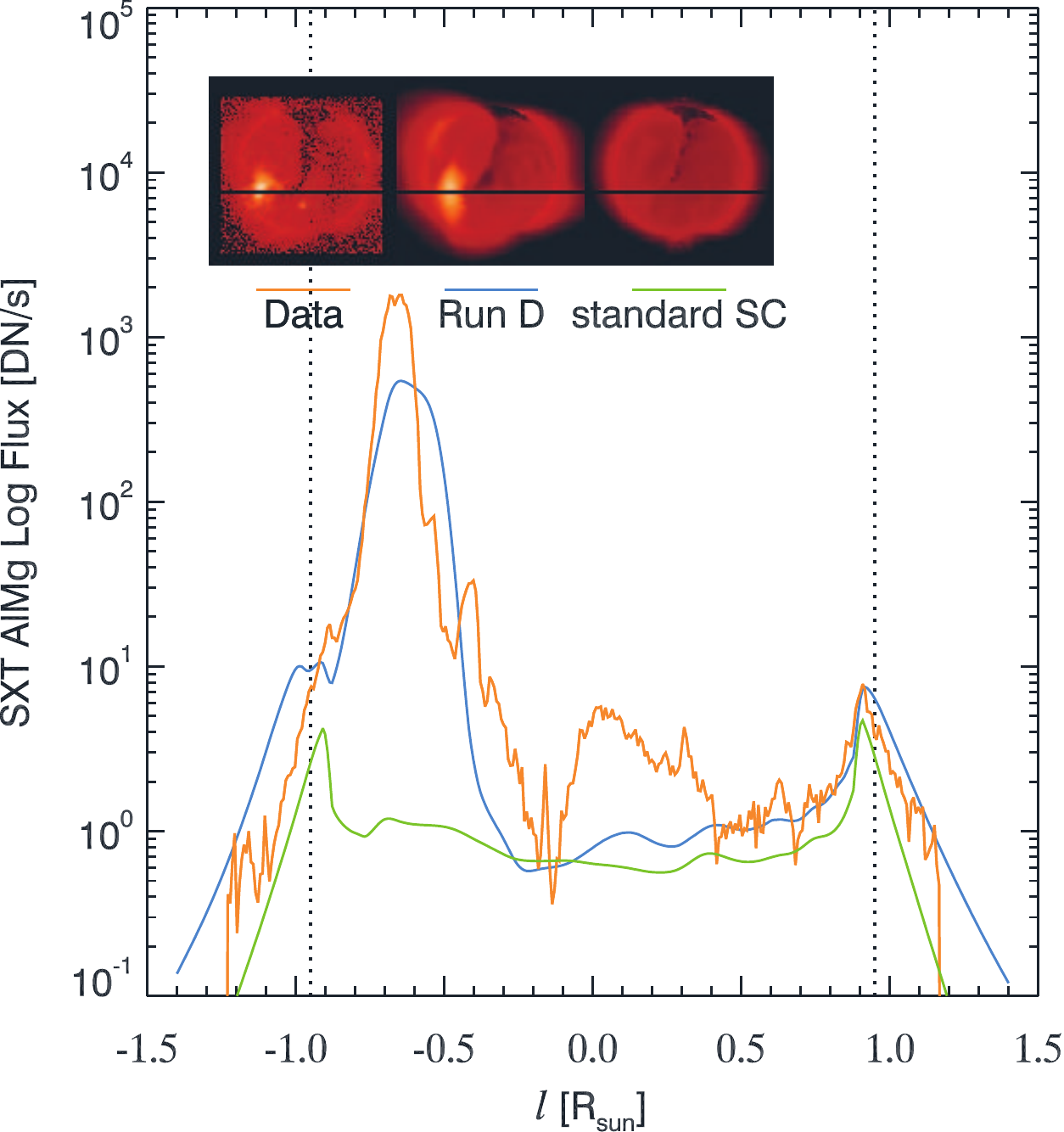}
\caption{\small Quantitative slice comparison of EIT 171\AA, 195\AA, 284\AA, and SXT AlMg synthesis from Run D (blue), the standard SC model (green), and observations (orange). The slices are chosen to include the average quiet Sun (left) and a large active region (right) observed on the disk.}
\label{fig:eit_slice} 
\end{figure}

In Figure \ref{fig:EUVcomp_run10toWSA} we show LOS EUV and Soft-X-Ray image synthesis comparisons between run D and the standard SC model. In Figure \ref{fig:eit_slice} we show two quantitative slice comparisons cutting an average quiet sun region and the large active region. The improved agreement of Run D in multiple filter bands is indicative of the improved temperature resolution possible with the LC model. While not an entirely fair comparison because the standard SC model does not include a full energy equation, we demonstrate the ability to quantitatively asses the relative agreement between a given model and observations in the EUV regime. This sort of comparison also allows one to study in detail various choices of heating models and, though multi-filter comparison, provide a basis to resolve the inherent degeneracies between modifying the total coronal heating power and the heating scale height within the model. A further examination of some of the limitations imposed by EUV synthesis from MHD models with finite resolution as well as future considerations can be found in Appendix \ref{section:appendix}.

\subsection{Thermodynamic Comparison}

\begin{figure}[hbtp]
\flushleft
\includegraphics[width=0.44\textwidth]{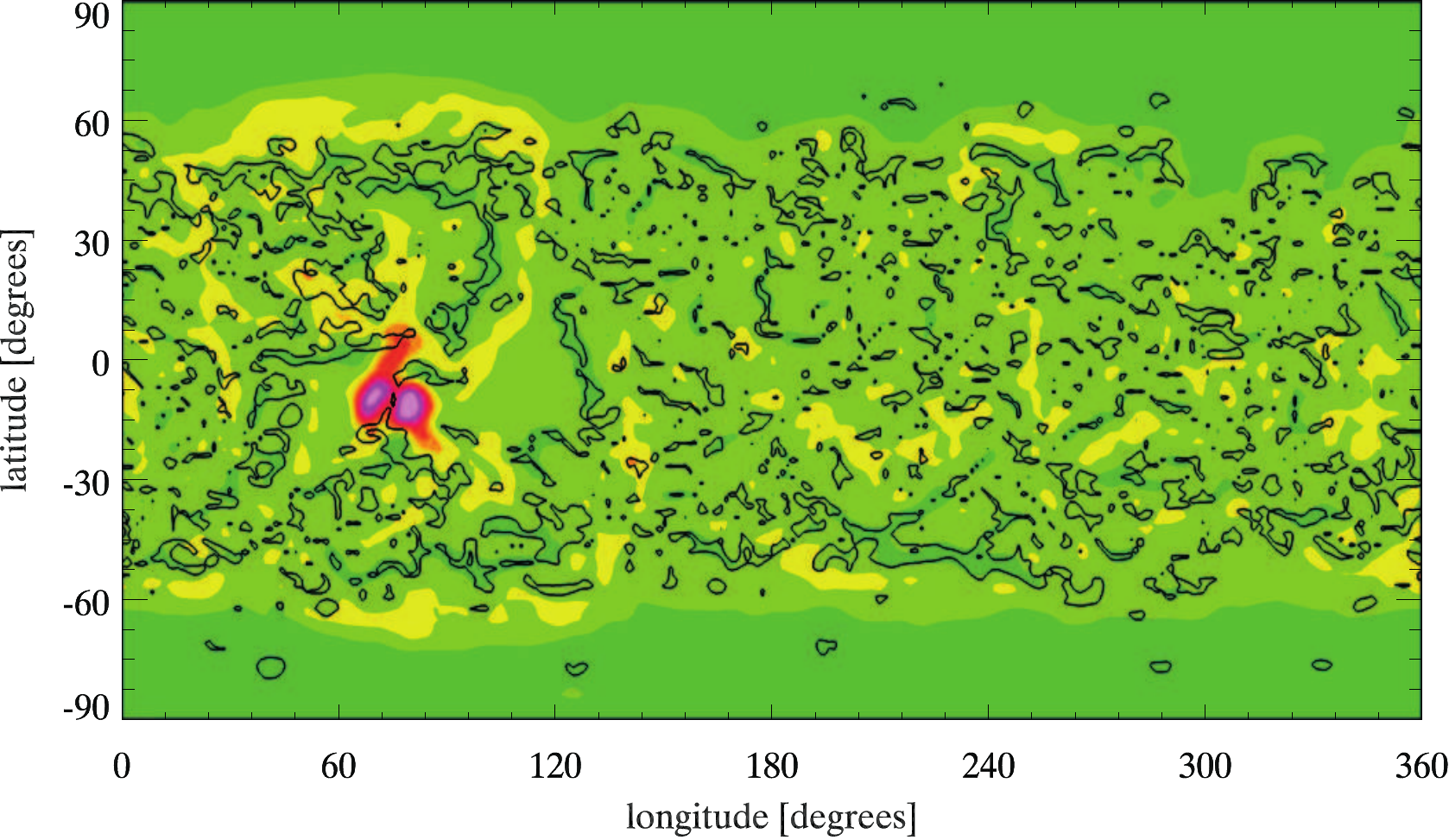}
\includegraphics[width=0.44\textwidth]{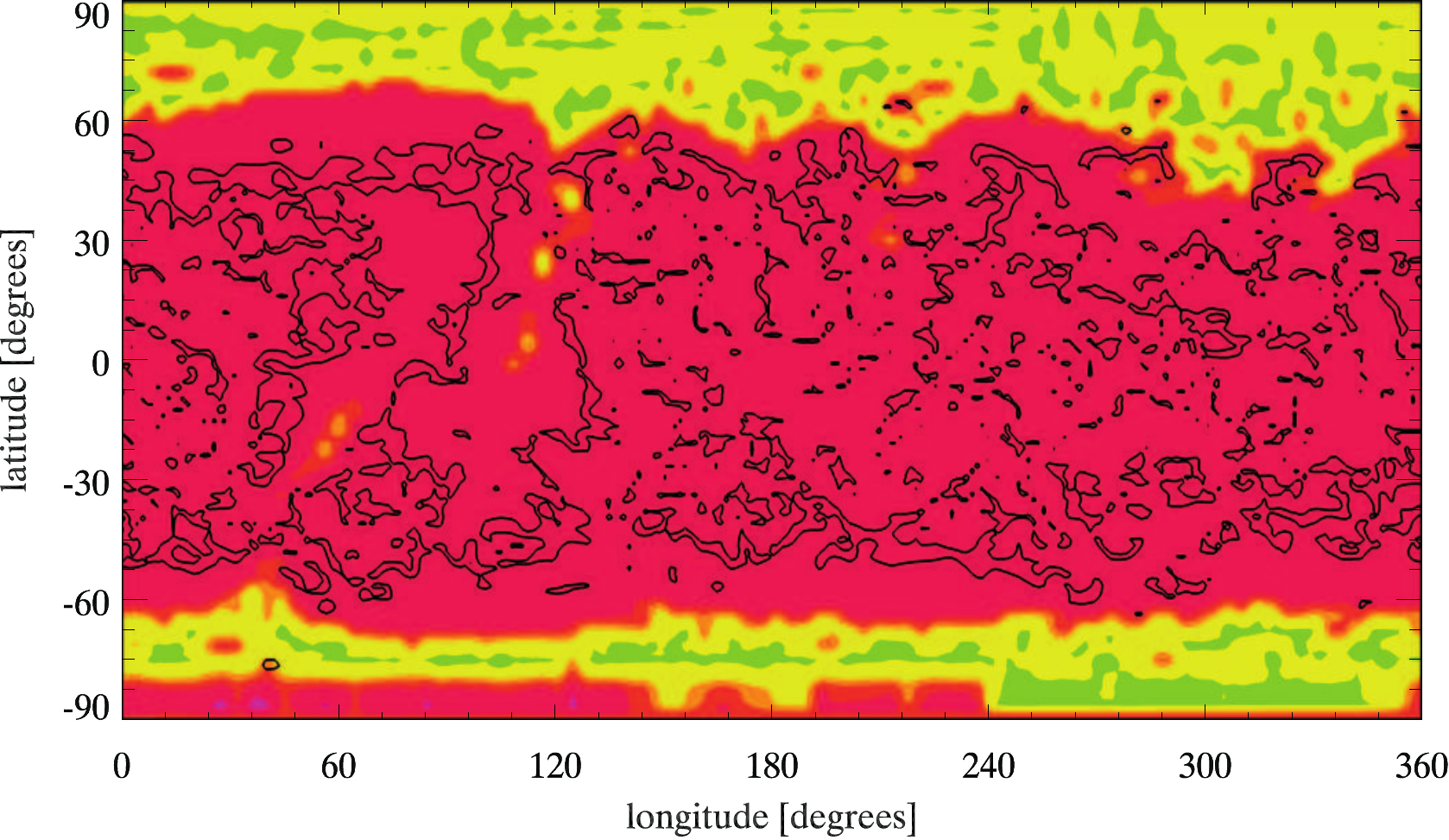}
\includegraphics[width=0.44\textwidth]{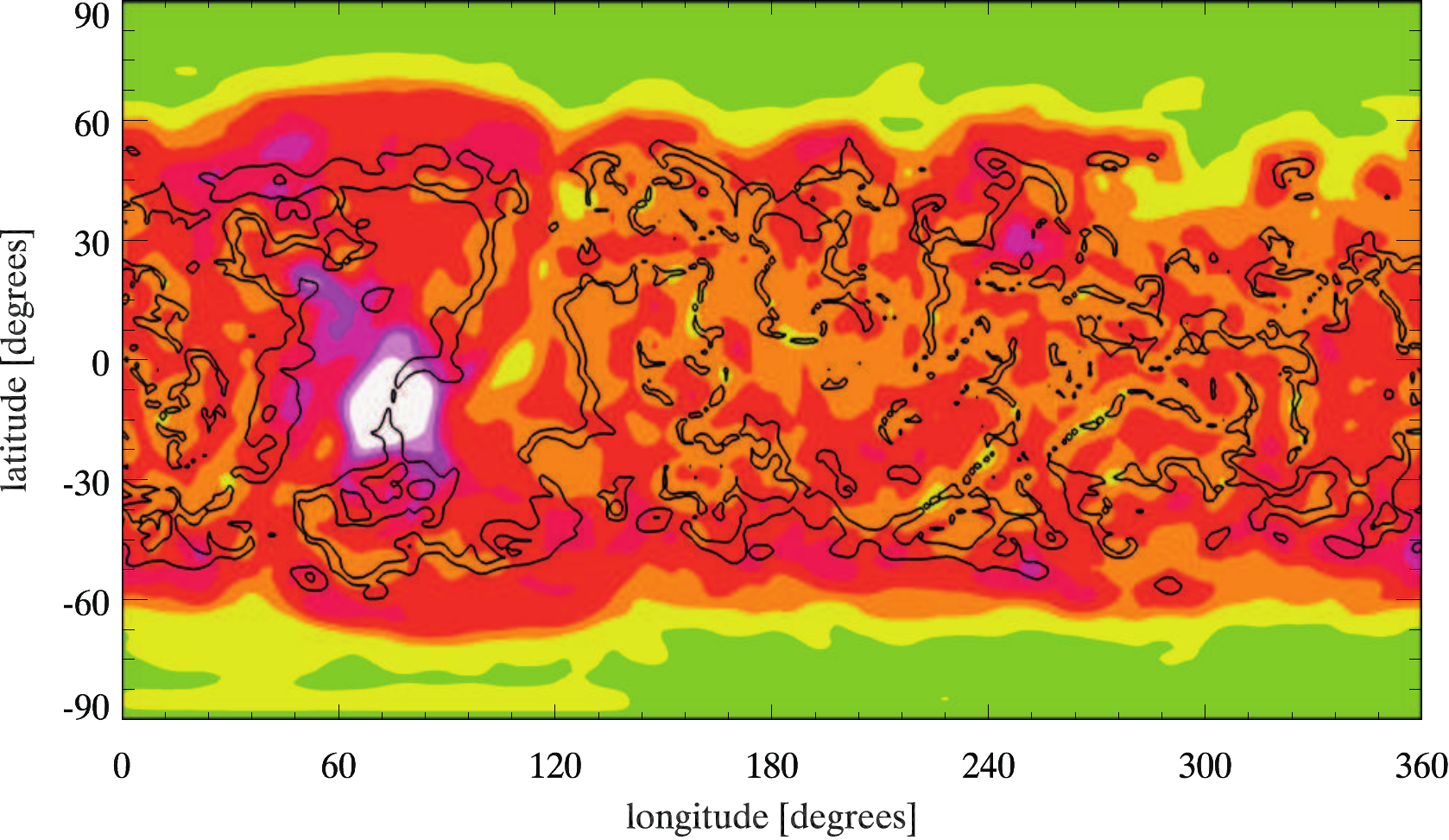}
\includegraphics[width=0.44\textwidth]{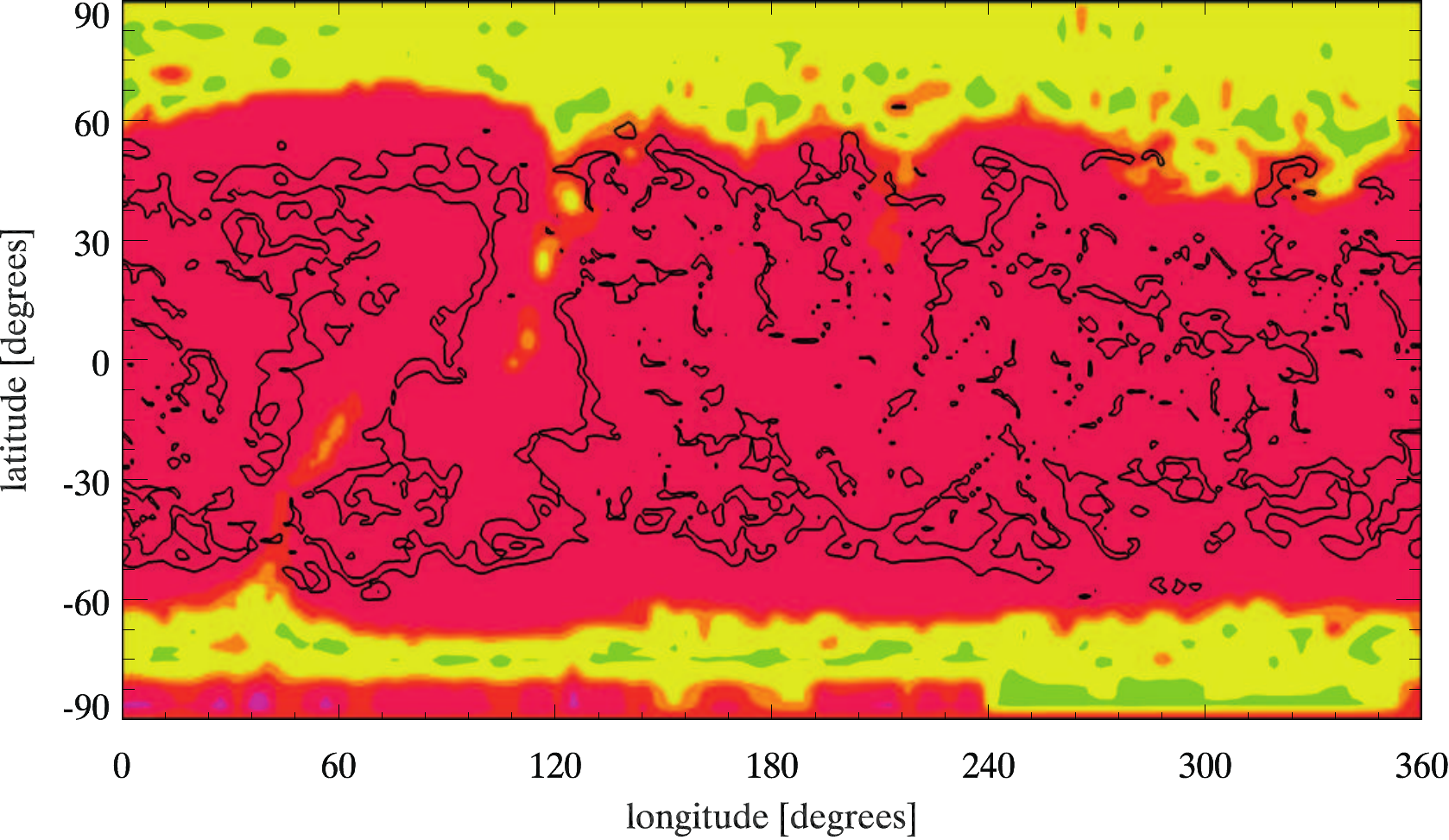}
\includegraphics[width=0.44\textwidth]{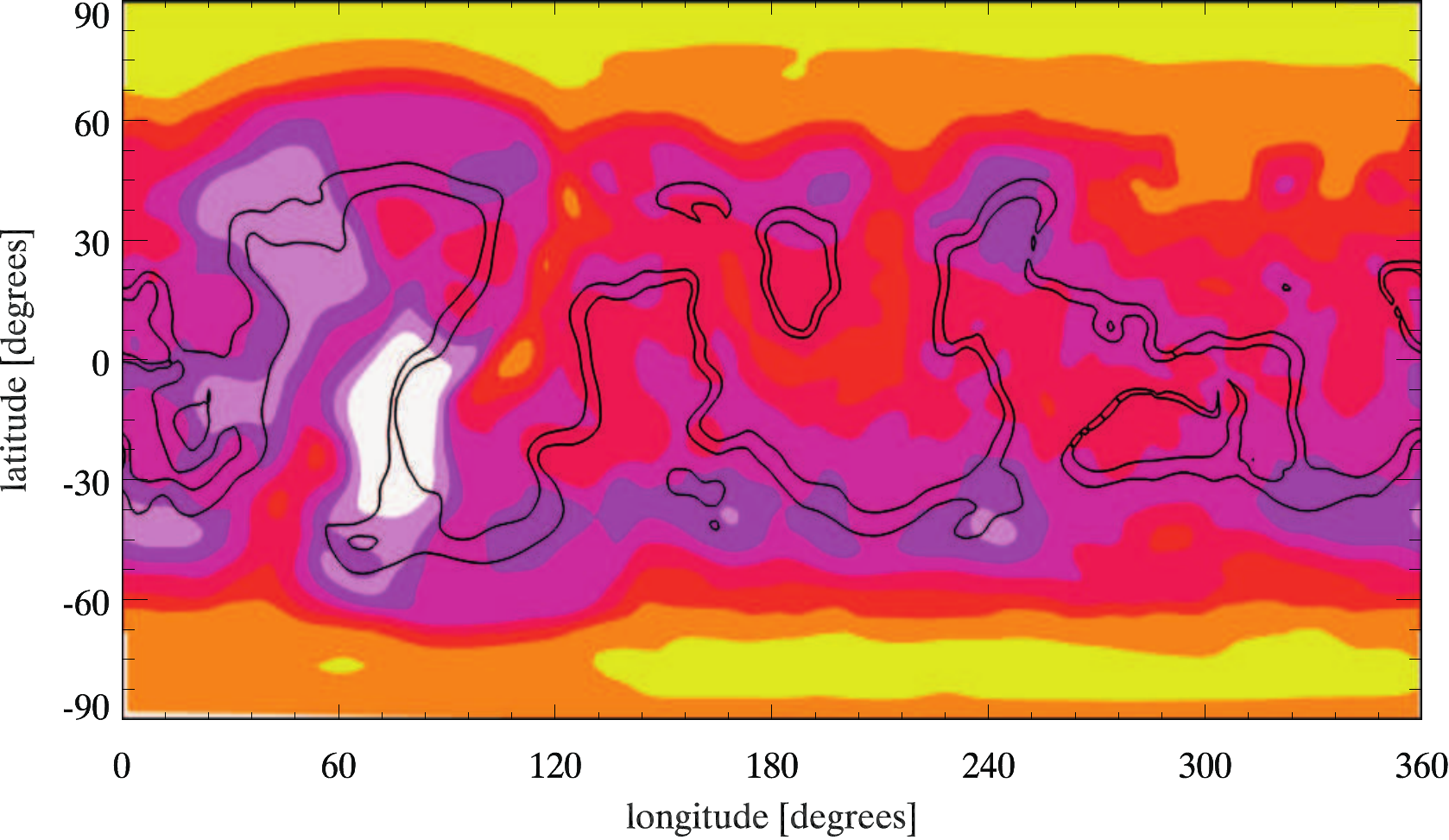}
\includegraphics[width=0.44\textwidth]{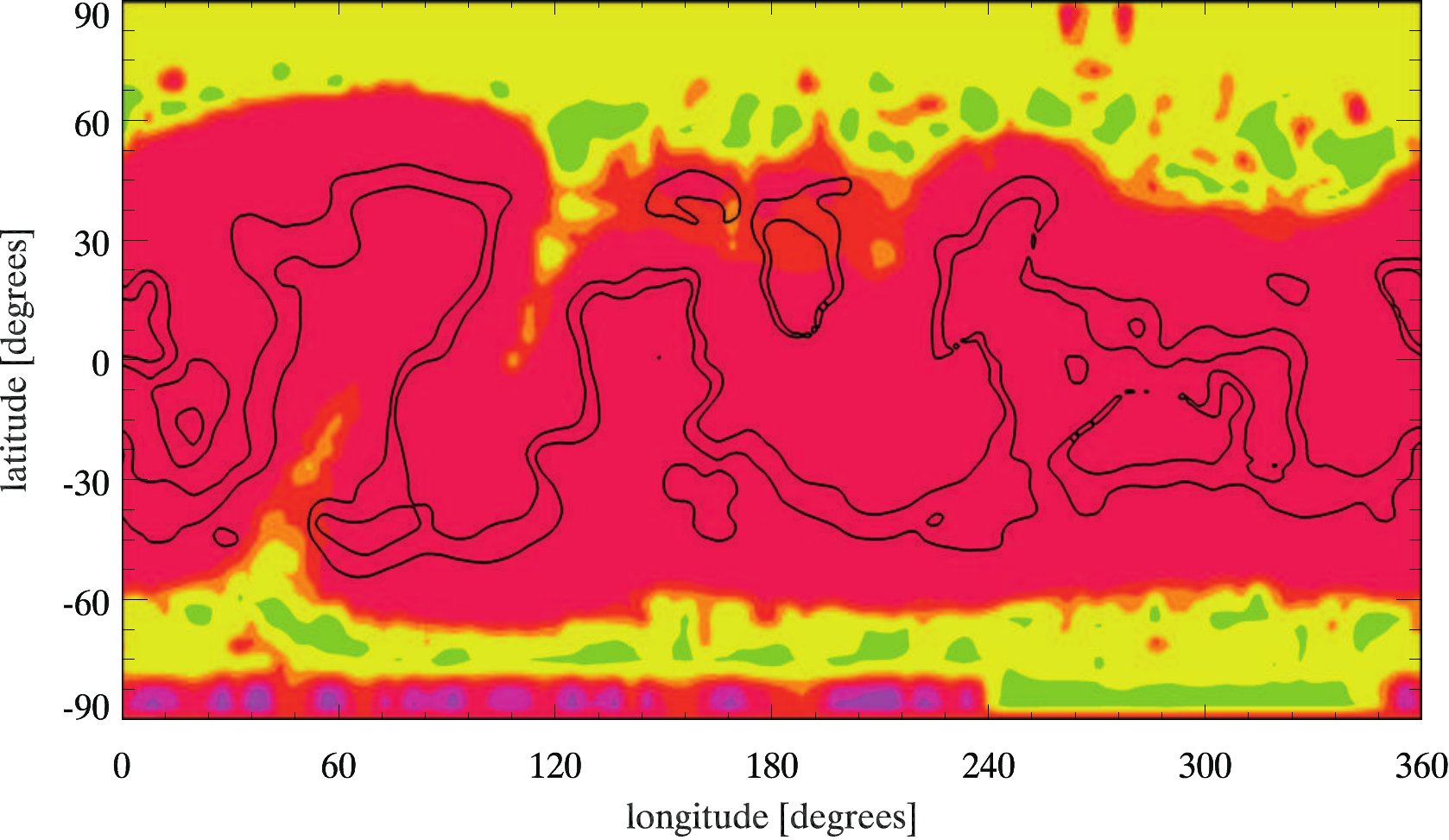}
\includegraphics[width=0.07\textwidth]{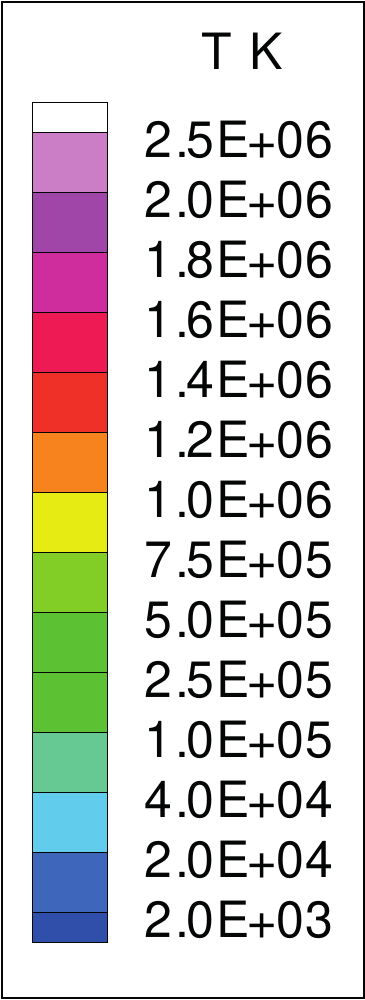}

\caption{\small Temperature contours at $r=1.01$ (top), $1.03$ (middle), and $1.10$ (bottom)$\ \text R_\sun$ for run D (left) and the default SC model (right) for CR 1913 (Black lines at $\hat{|B_r|}=0.2$). The full thermodynamic energy equation is able to capture large changes in temperature over short changes in radius. As expected, near the top of the transition region we resolve cooler temperatures near the inversion lines due the short loop length. Simultaneously higher in the corona, hotter temperatures ($T > 1.5 \text{MK}$) are achieved in close field regions, a natural result of energy flowing on a closed path via heat conduction.}
\label{fig:slices_Temperature} 
\end{figure}

\begin{figure}[hbtp]
\flushleft
\includegraphics[width=0.44\textwidth]{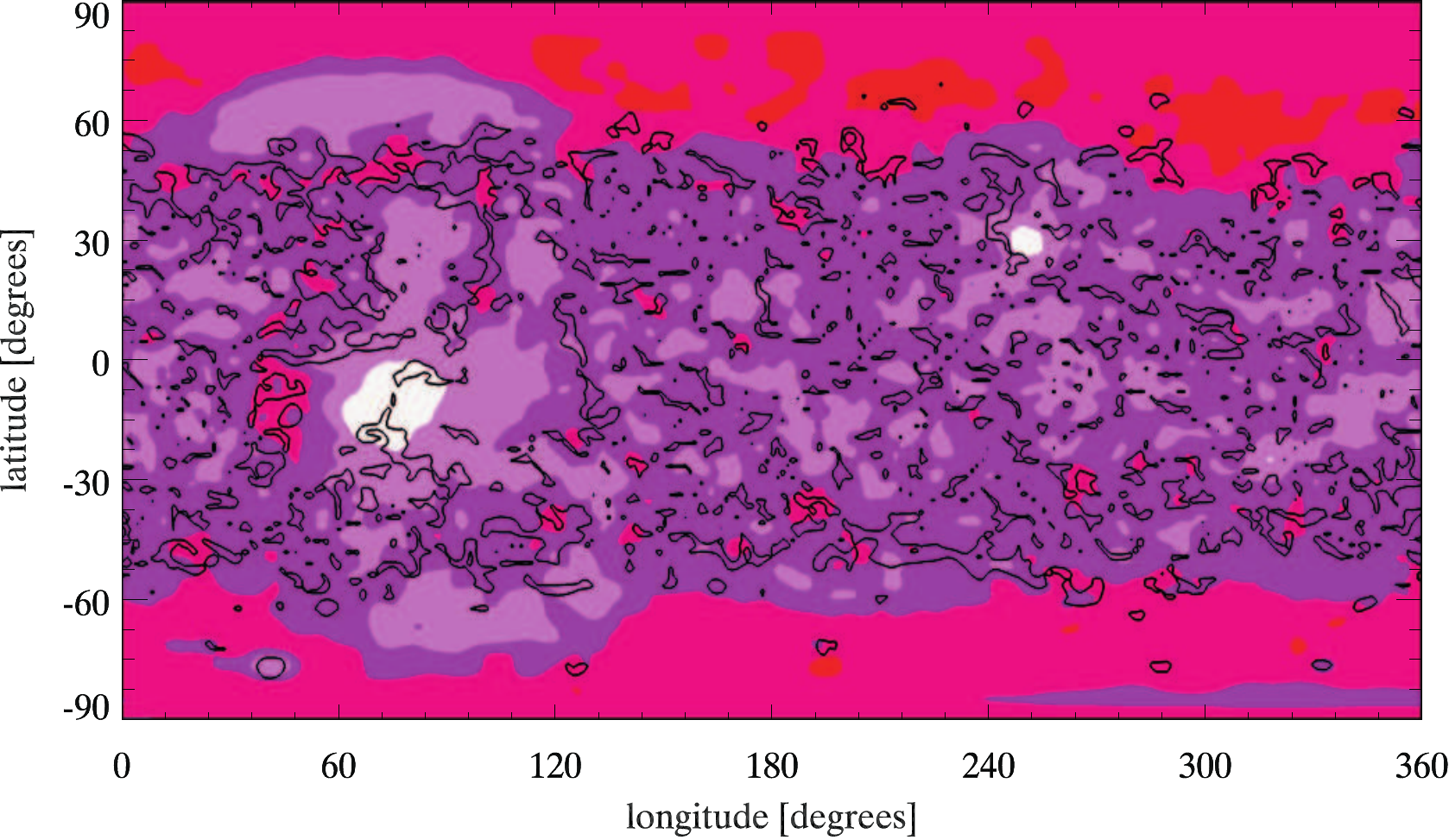}
\includegraphics[width=0.44\textwidth]{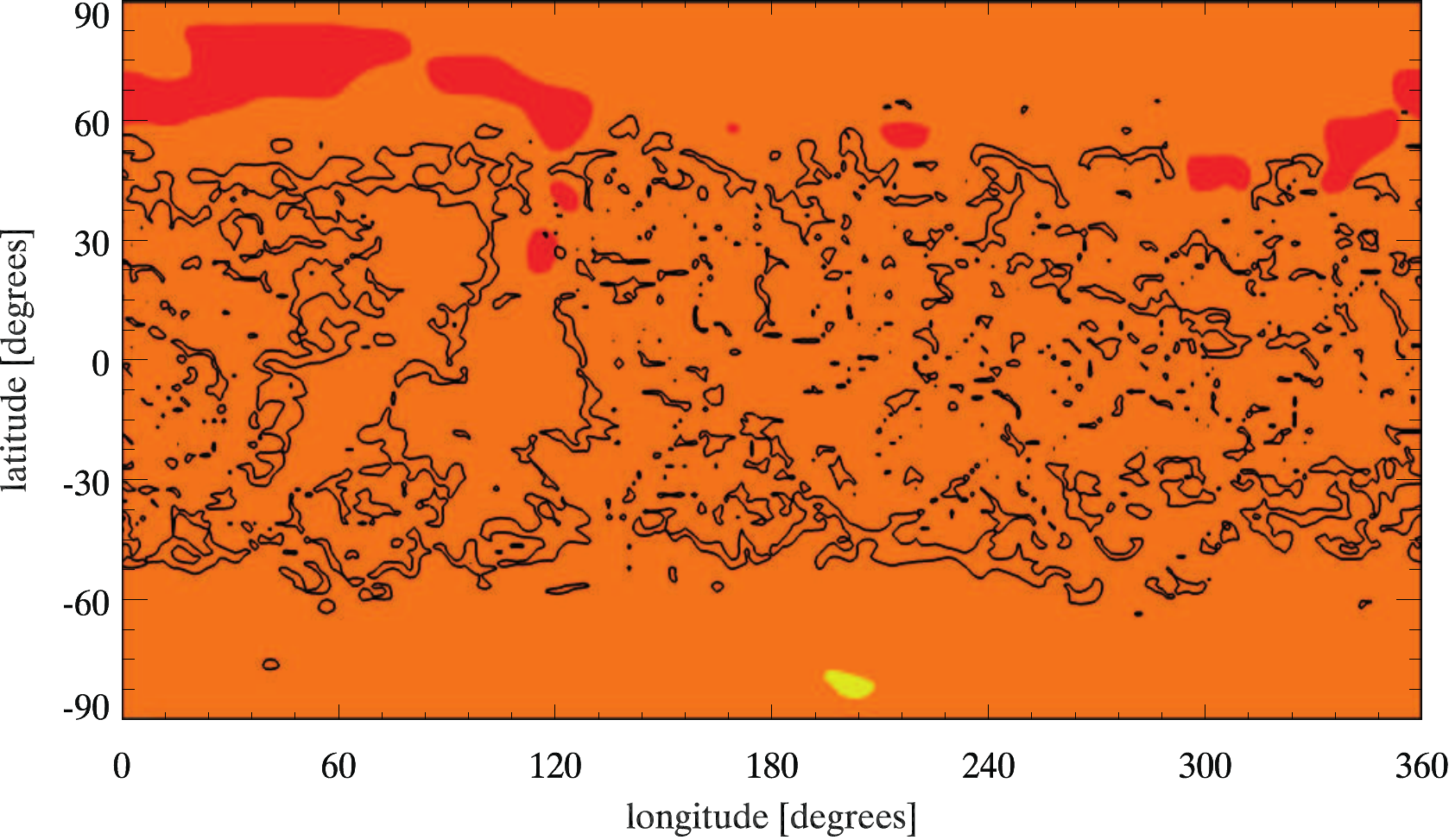}
\includegraphics[width=0.44\textwidth]{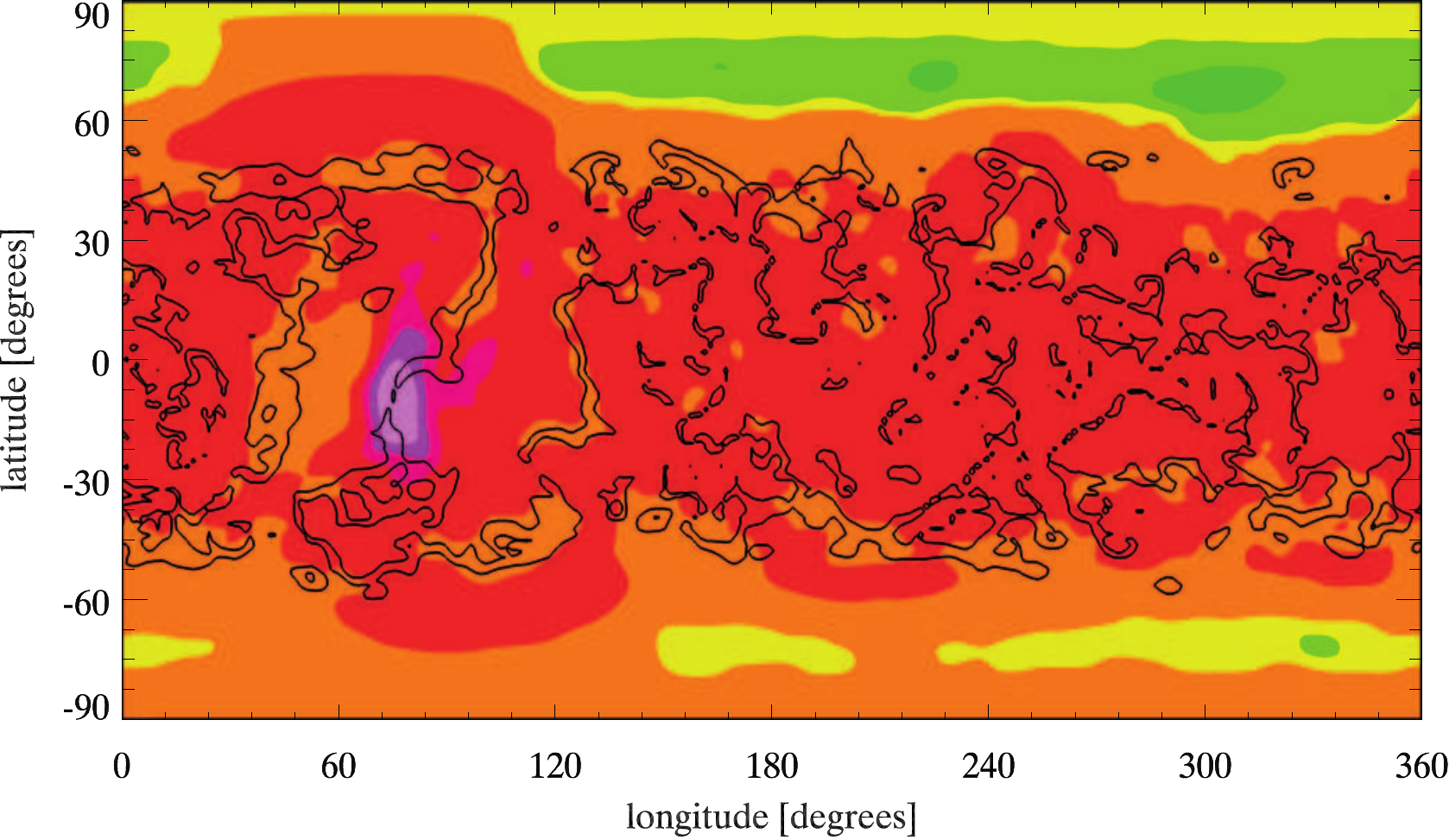}
\includegraphics[width=0.44\textwidth]{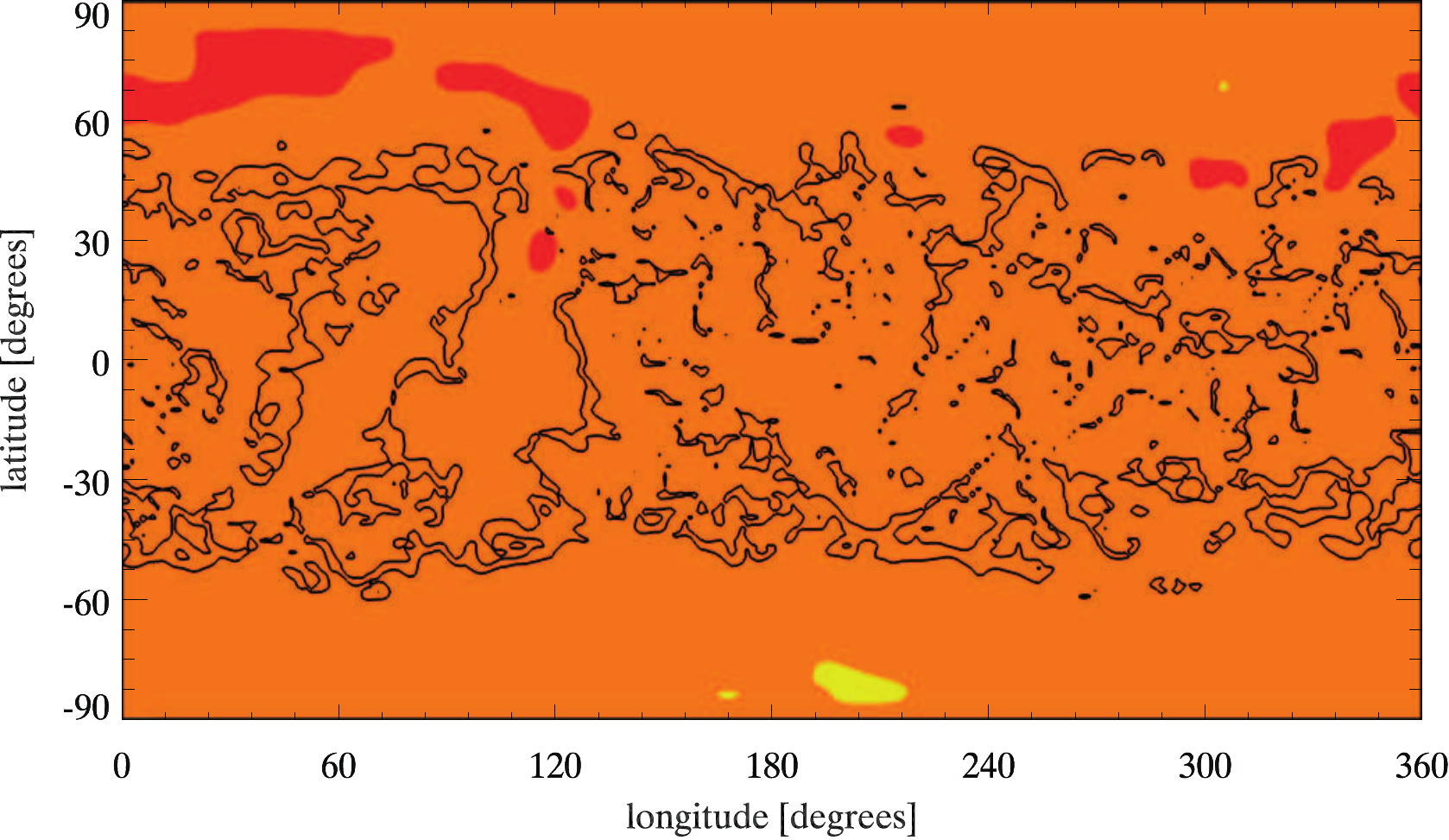}
\includegraphics[width=0.44\textwidth]{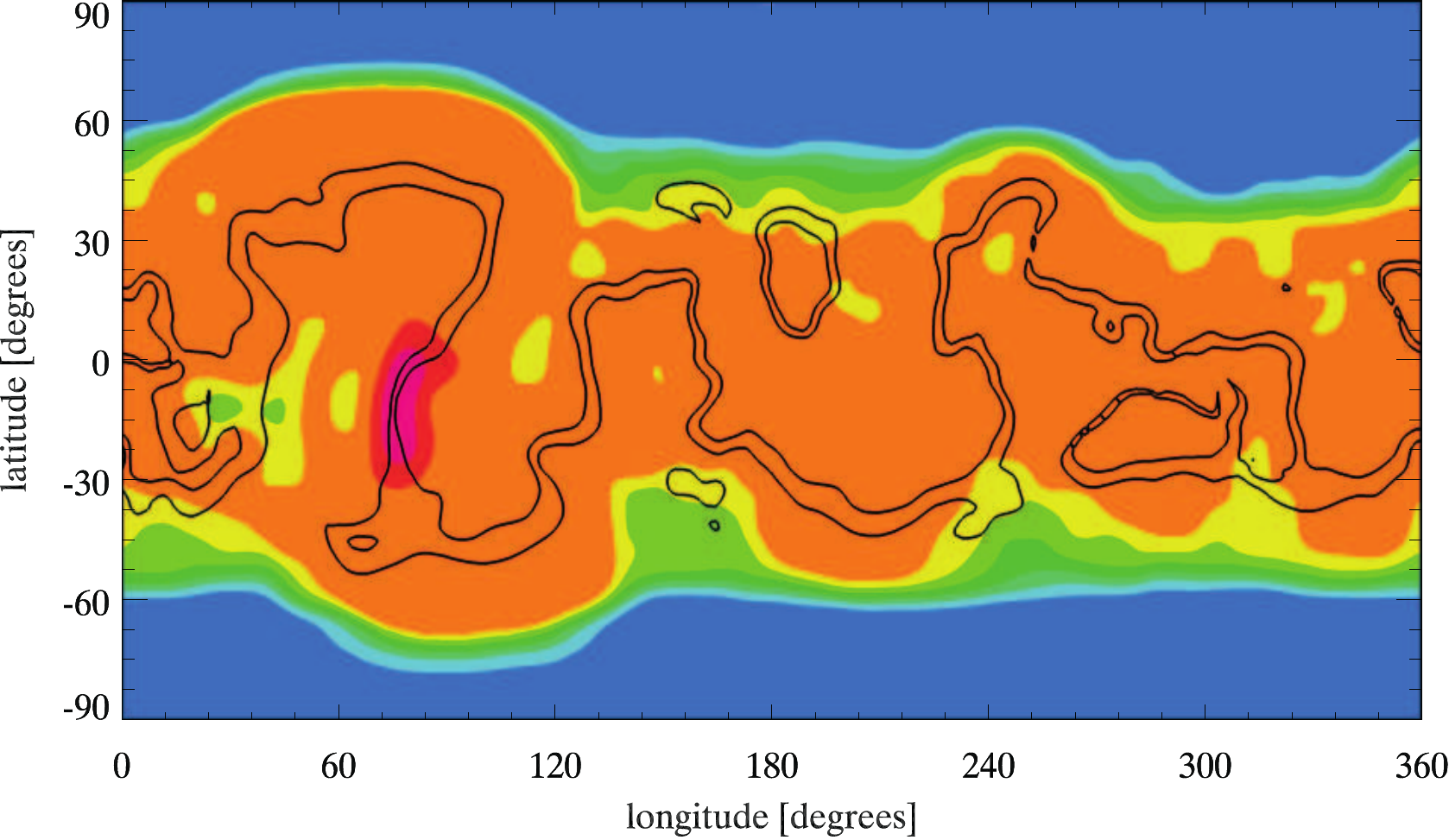}
\includegraphics[width=0.44\textwidth]{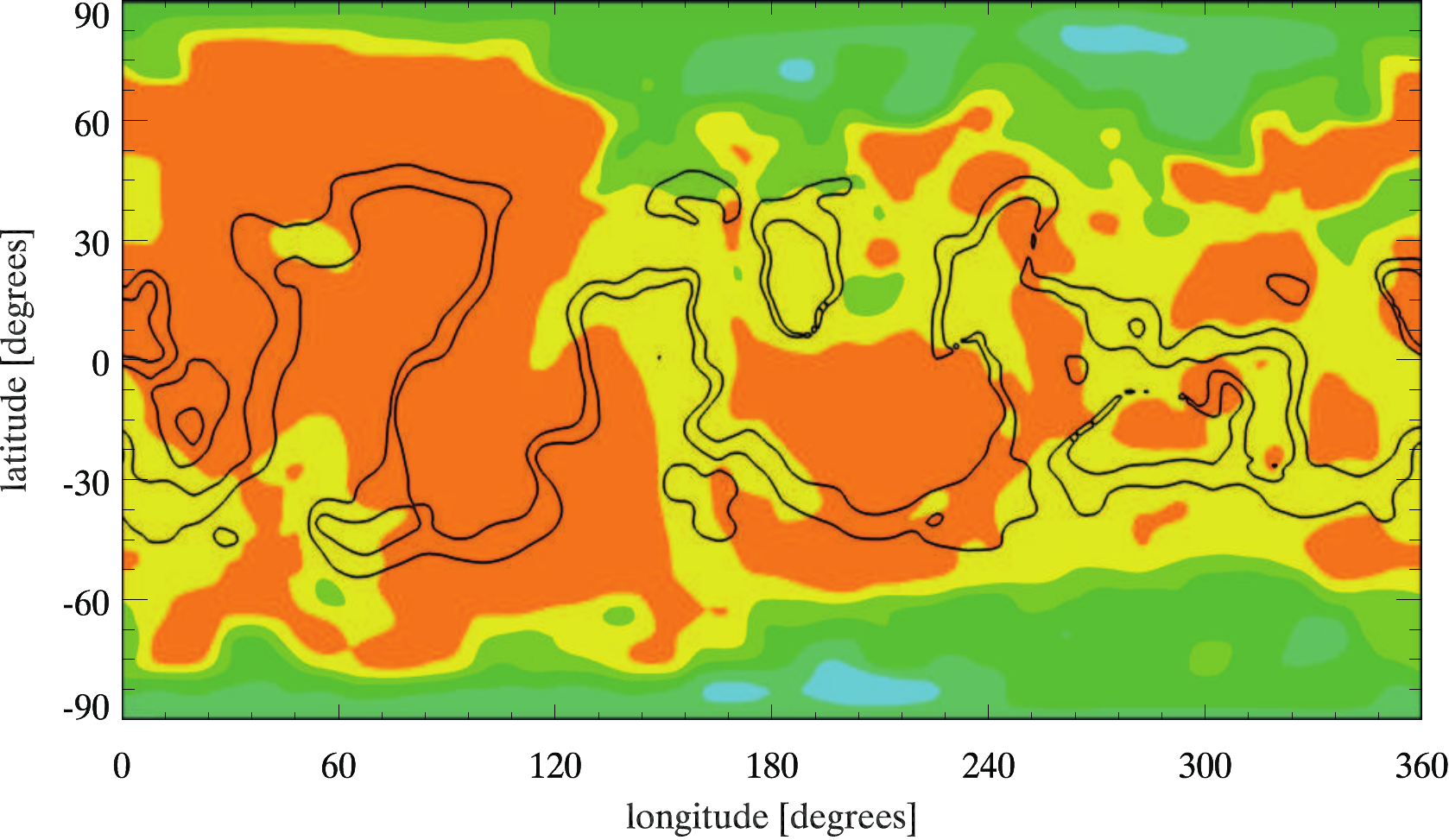}
\includegraphics[width=0.07\textwidth]{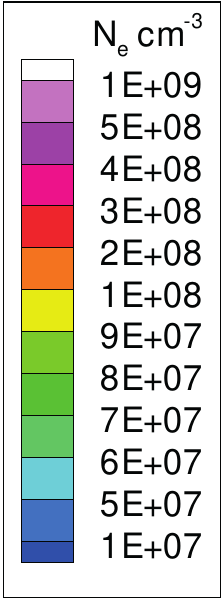}

\caption{\small Electron density contours at $r=1.01$ (top), $1.03$ (middle), and $1.10$ (bottom)$\ \text R_\sun$ for run D (left) and the standard SC model (right) for CR 1913 (Black lines at $\hat{|B_r|}=0.2$). This demonstrates the high dynamic range in density over small changes in radius resolved with the LC model.}
\label{fig:slices_density} 
\end{figure}

It is also instructive to examine the global temperature and density structure as a function of height for both Run D and the standard SC model. In Figures \ref{fig:slices_Temperature} and \ref{fig:slices_density} we show electron temperature and number density at three separate heights in the low corona. Important to emphasize is the fact that with a thermodynamic model, it is possible to resolve both significant changes in temperature with height and with respect to the magnetic field topology in the model. As expected, near the top of the transition region Run D resolves cooler temperatures near the inversion lines due the short loop length, while simultaneously higher in the corona, hotter temperatures ($T > 1.5 \text{MK}$) are achieved in close field regions, a natural result of energy flowing on a closed path via heat conduction. This greater sensitivity to field geometry (something difficult for a polytropic model to do without extreme fine tuning of boundary conditions) is critical when trying to correctly describe the conditions in the low corona. 
\\

\section{Conclusion}
\label{section:conclusion}

In this paper we demonstrate the clear need for a self-consistent treatment of the thermodynamic energy equation and boundary conditions when studying the global properties of the low corona with an MHD model. We observe that the interplay between coronal heating and electron heat conduction strongly governs the details of important strucures in the low corona, which is something that connot be described by extrapolation of the magnetic field on its own (e.g. the significant stress of the streamer regions and current sheet in response to active region heating, Figures \ref{fig:streamer_Br_slice} and \ref{fig:Magnetic_topology}). We also showcase the ability of the LC model to examine the effectiveness and applicability of various heating functions, and find that a simple exponential heating profile does quite well in describing the average structure of the quiet sun.
\\

With this foundation, we believe that the LC model can provide a more realistic vehicle with which to study a wide range of aspects of the dynamic corona and solar wind. For example, the time-dependent dynamics of any sort of transient event in the corona, such as a CME or EUV wave, depend critically on the magnetic and density structure of the corona (density scale height being highly sensitive to temperature) and therefore it is important to use improved models when determining the global ambient medium in which they evolve. Additionally because of the general flexibility of the SWMF, future coupling of the LC  model to complementary solar wind models, such as the SC model based on a full description of Alfv\'en wave turbulence currently in development \citep[in press]{oran09}, or one including a multi-species treatement to resolve electron/ion temperature decoupling in the high corona, has become a feasible scenario. Ultimately, one thing is clear: with each passing month, year, and decade, our understanding of these fundamental process is constantly refined. In furthering this process it is critical to continue to improve the existing models and validate new efforts as they emerge.

\acknowledgements

This work is primarily supported under the NASA Earth and Space Science Fellowship (NESSF) Program NASA NESSF08-Helio08F-0007 as well as grants: NSF ATM-0639335 (CAREER) and NASA NNX08AQ16G (LWS). Computations were performed on the computing cluster at the Institute for Astronomy (IFA) Advanced Technology Research Center (ATRC) Pukalani, HI. C. Downs would like to thank the entire SWMF team for their innovative research and willing collaboration. 
\\

SOHO EIT images courtesy of The SOHO EIT Consortium; SOHO is a project of international cooperation between ESA and NASA. Yohkoh data courtesy of the NASA-supported Yohkoh Legacy Archive at Montana State University.

\appendix
\section{Interpreting synthesized EUV images from global models}
\label{section:appendix}

One critical caveat when imaging the solar corona is that remote sensing images are fundamentally ambiguous in the sense that they are 2D realizations of a fully 3D, optically thin data set. For our purposes, it is best to frame the problem with the concepts of modeling an unresolved Differential Emission Measure (DEM) distribution and Hydrostatic Weighting Bias, well described in \citet{aschwanden00,aschwanden01}. These issues arise due to two main facts. (i) The spatial resolution of current imaging instruments (e.g. TRACE, SOHO, STEREO) all lie below the fundamental width of an isolated flow/field line. And (ii) any image is a line of sight integral over a range of heights, which are preferentially biased towards hotter flow lines with longer density scale heights. Thus at further distance off of the limb, the observation may be biased towards flow lines that may not be energetically important at the surface. Local DEM distributions can also be directly observed through coupled filter and tomographic inversions \citep{frazin09}.
\\

Though the LOS integral for image synthesis \eqref{los} includes the range of temperatures and densities in cells along the line of sight, it cannot fundamentally account for the DEM distribution that is below the physical resolution of the model. In MHD, a given cell has a unique value for its thermodynamic properties, particularly $n_e$, and $T$, while a physical picture of the low $\beta$ multi-hydrostatic corona below the resolution limit can be treated as a superposition of densities and temperatures, each confined to an isothermal flow/field line. In this sense, the true DEM distribution of a parcel of gas that contributes to an EUV image (e.g. the EIT instrument) is instead represented as a delta function in the simulation. One aspect of this limitation can be seen in difference between the data and synthesized limb profiles (Figure \ref{fig:eit_slice}). Because the observational weighting bias shifts to higher temperatures further with distance from the surface, the observed limb profile cannot be represented in all four filters simultaneously with only the larger scale temperature equilibrium produced by the MHD model.
\\

\begin{figure}[hbtp]
\centering
\includegraphics[width=0.485\textwidth]{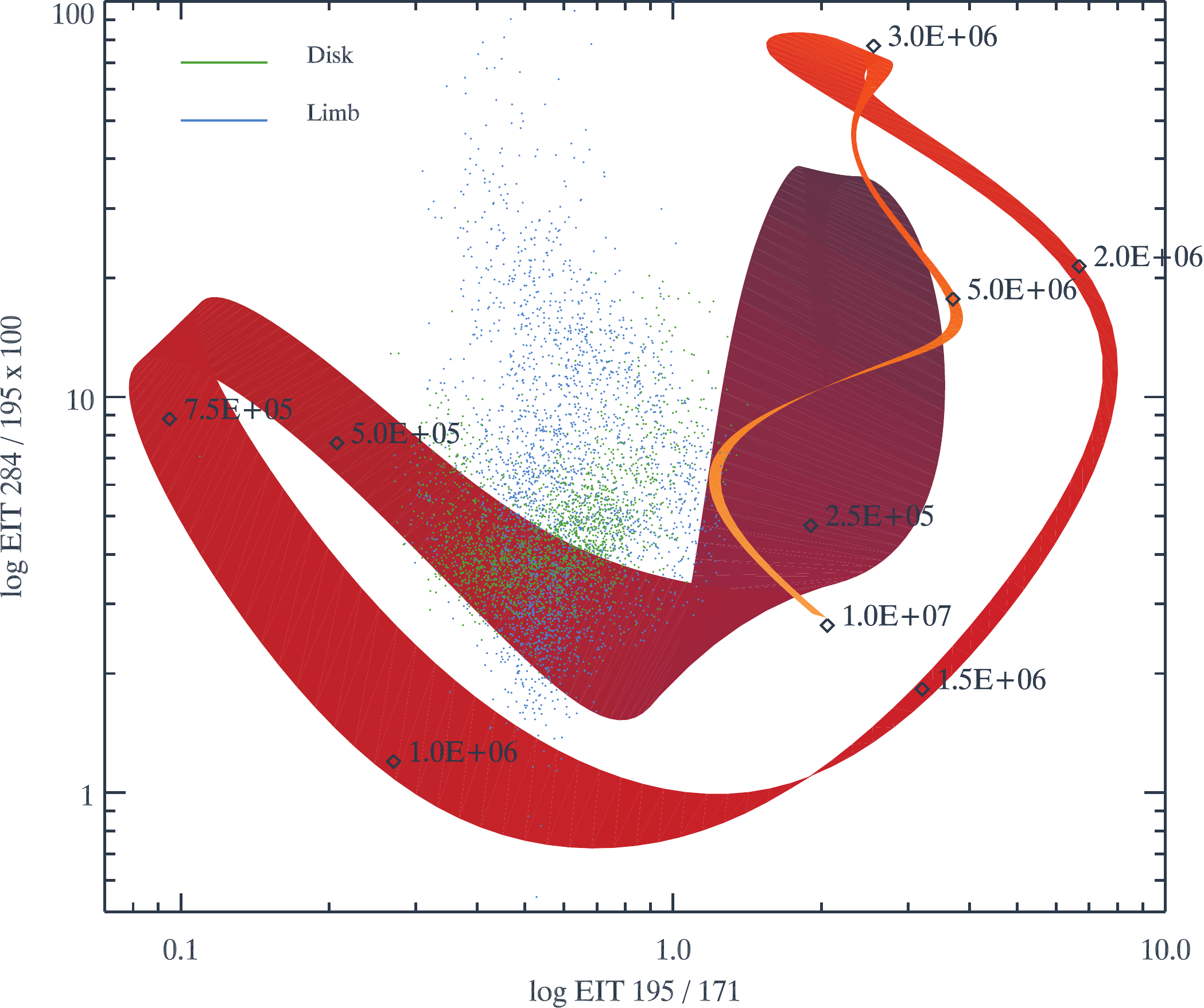}
\includegraphics[width=0.49\textwidth]{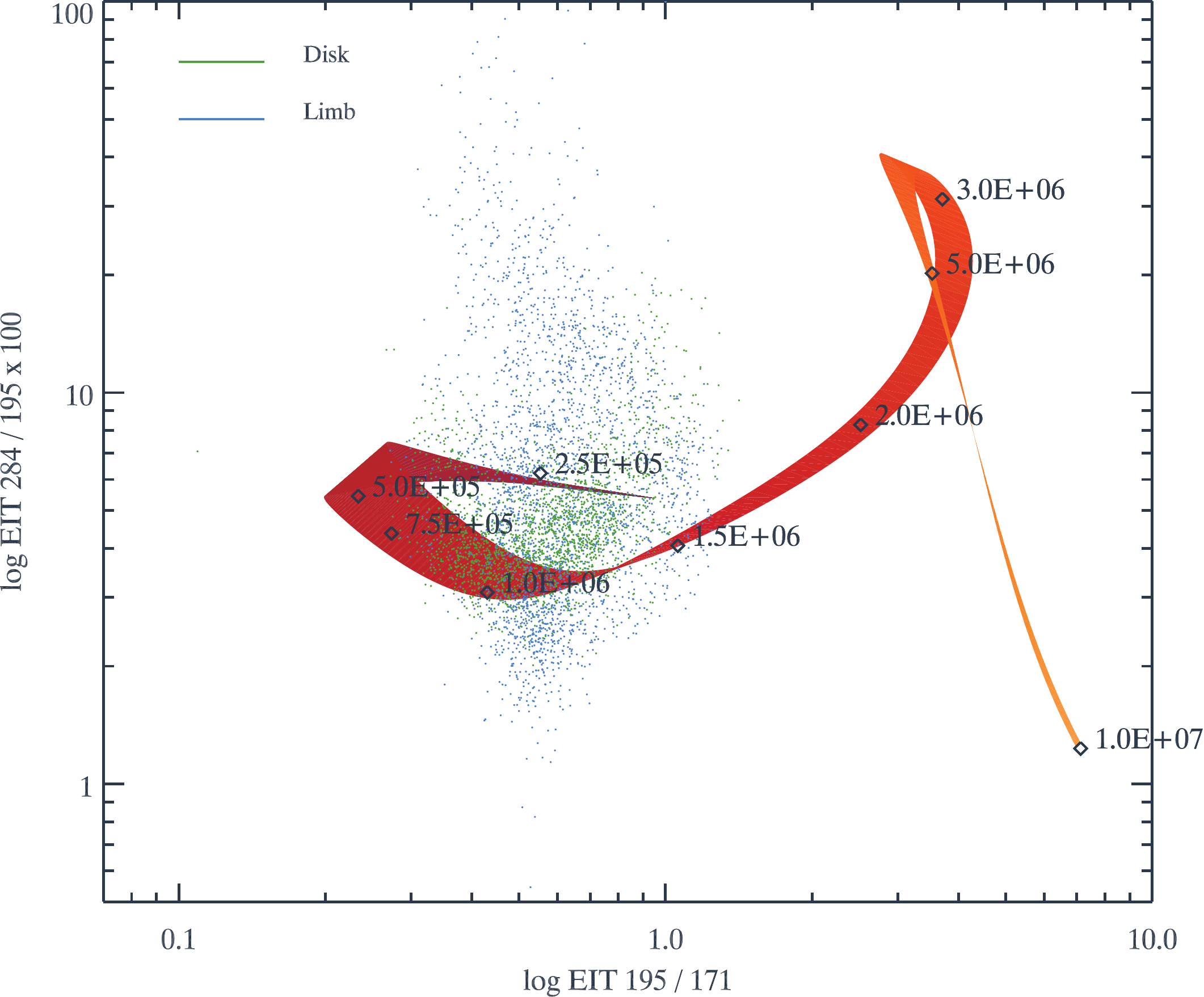}
\caption{\small EIT filter ratio path with no DEM modification (left) and 
ratio path with Gaussian convolution of response with $\sigma_T = 0.5 T_0$ (right). The blue and green points represent EUV data on an off the limb (binned to 1000 points each). The curve path follows the ratio values [$f_{195}(T,n_e)/f_{171}(T,n_e)$ and $f_{284}(T,n_e)/f_{195}(T,n_e)]$ from $T=2.5 \times 10^{5} - 10^7$K, the width of the curves represents the variation of the ratio values from $n_e = 1 \times 10^{6-12}\text{cm}^{-3}$ at a given temperature. This is intended to illustrate the large variation of filter ratio values with respect to an unresolved DEM distribution.}
\label{fig:DEMcurves} 
\end{figure}

To further examine this issue in the context of the LC model we use filter ratios between the EUV images, which are nominally sensitive to temperature differences. In Figure \ref{fig:DEMcurves} we show filter ratio values for SOHO EIT observations on Aug 27, 1996 and the path in the filter ratio plane spanned by all possible filter ratios of the optically thin response functions $f_{195}(T,n_e)/f_{171}(T,n_e)$ and $f_{284}(T,n_e)/f_{195}(T,n_e)$ (section \ref{section:LOS-synthesis}). Unsurprisingly, the EUV data spans a much larger range than that allowed by a single temperature. To simply examine the effect that an unresolved DEM distribution may have on the synthesis and subsequent comparison to observational data we construct an ad hoc DEM by Gaussian convolution of the filter response functions with T:
\begin{equation} \label{eq:DEM}
f^{mod}_{i}(T_0,n_e) = \frac{1}{\sqrt{\pi}\sigma_T}\int_{10^4}^{10^7}{f_{i}(T,n_e)e^{-\frac{{(T-T_0)}^2}{\sigma_T^2})}dT}
\end{equation}
and use $\sigma_T = 0.5T_0$ for this discussion. While we by no means posit that this DEM distribution should resemble the corona everywhere, it includes the basic temperature splitting features of those gaussian DEM distributions calculated from observations (e.g.  \citet{aschwanden01,frazin09}. We also show the effect of this modification in the right panel of Figure \ref{fig:DEMcurves}.
\\

\begin{figure}[hbtp]
\centering
\includegraphics[width=0.49\textwidth]{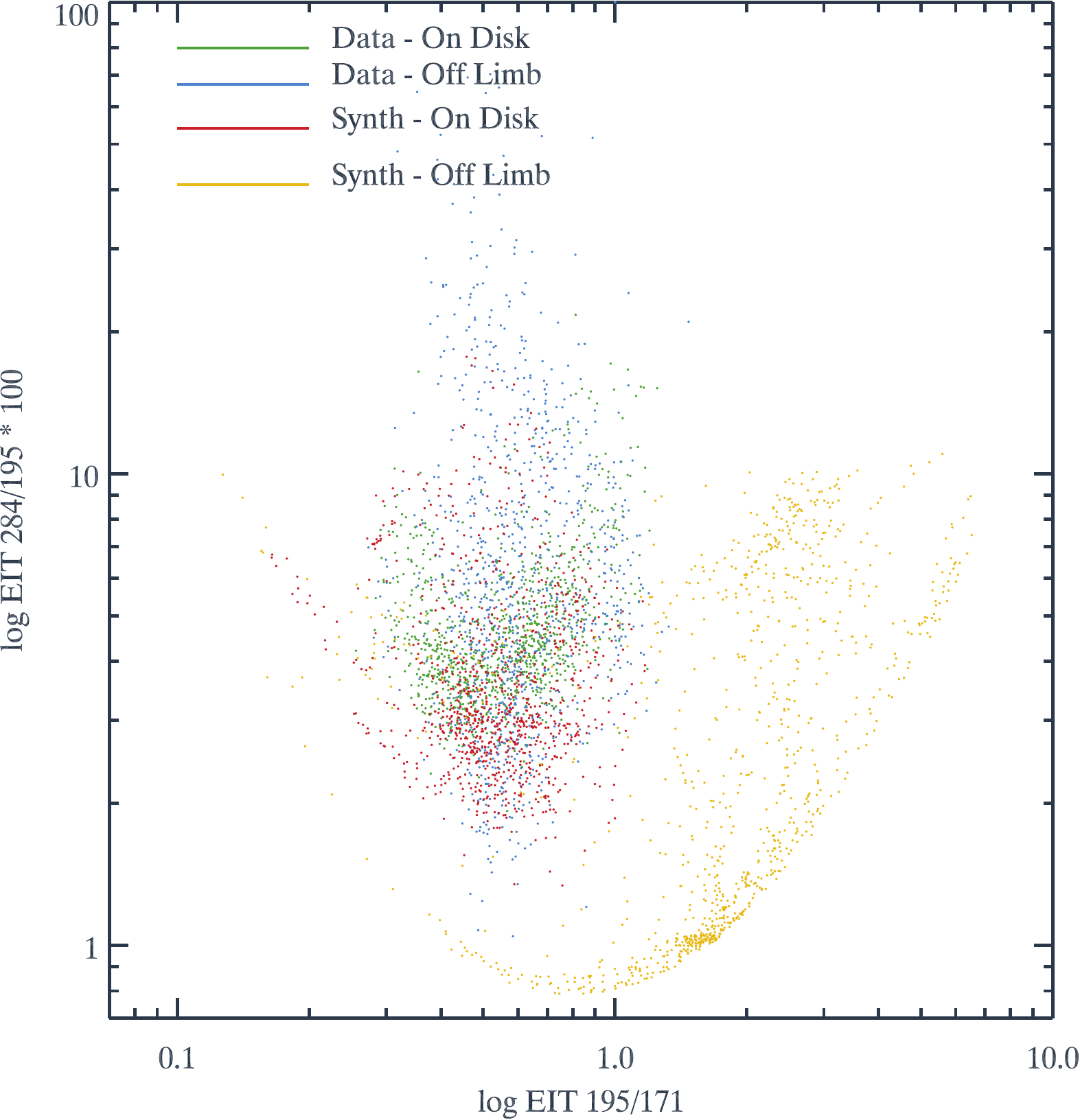}
\includegraphics[width=0.49\textwidth]{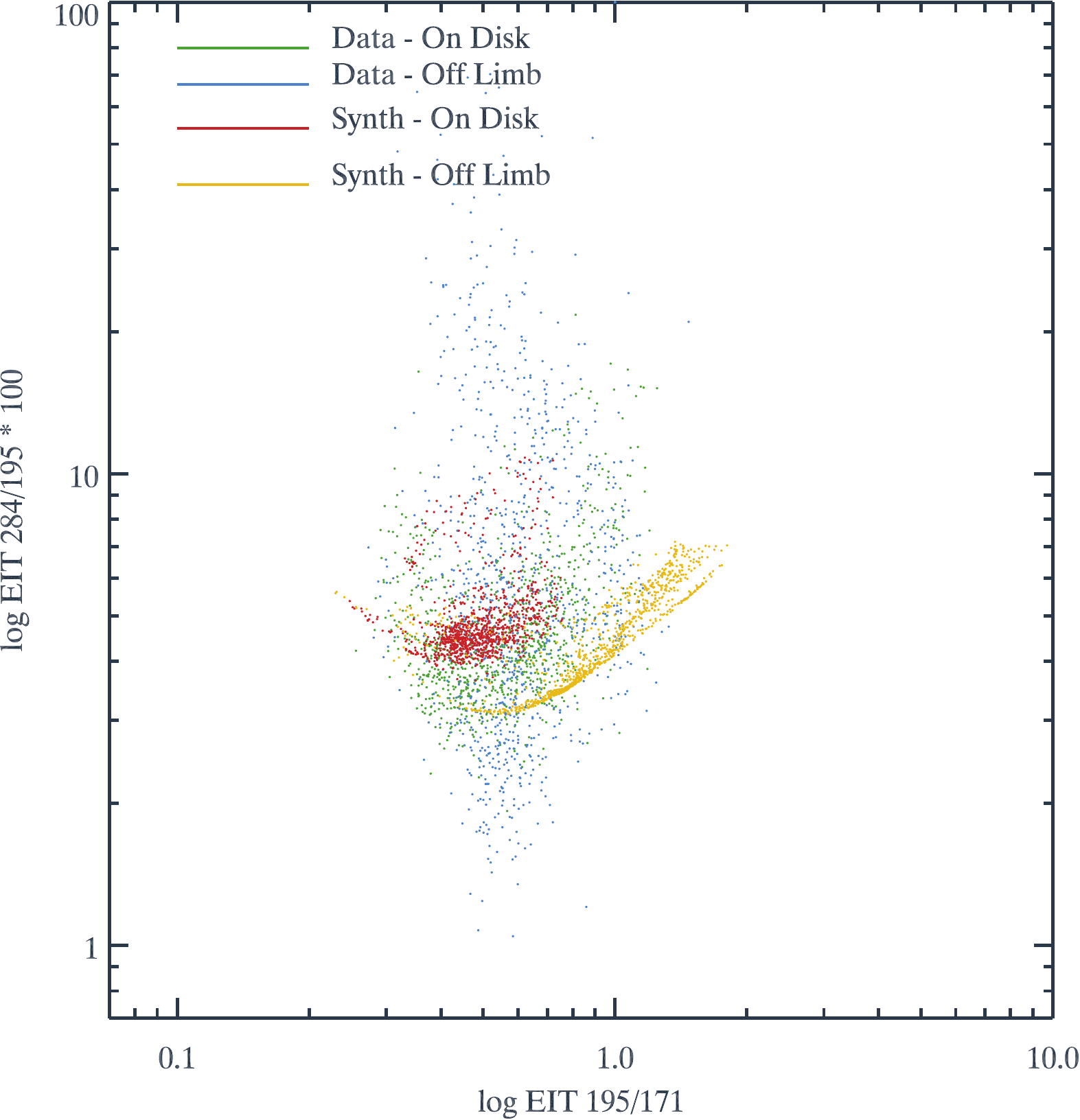}
\caption{\small Effect of simple DEM modification on the synthesized EIT filter ratios for Run D. LOS synthesis ratios with no DEM modification (left) and those with Gaussian convolution of response with $\sigma_T = 0.5 T_0$ (right). While the location of the synthesized filter ratios improves with this DEM modification, the fine details of the local DEM distribution, realized by the variation of observed EIT filter ratio values, are difficult to reproduce with an MHD model involving large, global scales.}
\label{fig:DEM_effect_obs} 
\end{figure}

The effect that this simple induced DEM distribution has on the synthesized ratios is shown in Figure \ref{fig:DEM_effect_obs}. It is clear comparing the limb tracks of the synthesized images to the temperature tracks in Figure \ref{fig:DEMcurves} that the LOS synthesis at the current model resolution does not produce a significant DEM distribution on its own (particularly at the limb). However, the improvement gained in terms of the relative location with respect to observations for the modified DEM synthesis suggests that this effect should not be overlooked when constructing more detailed models in the future. High resolution studies designed to approach the fundamental scale of isothermal separation between coronal loops, particularly those involving local, time-dependent heating for example, should be able to address this fundamental issue.


\clearpage

\section{Figures}


\clearpage


\begin{thebibliography}{41}
\expandafter\ifx\csname natexlab\endcsname\relax\def\natexlab#1{#1}\fi

\bibitem[{Abbett(2007)}]{abbett07}
Abbett, W.~P. 2007, \apj, 665, 1469

\bibitem[{Altschuler {et~al.}(1977)Altschuler, Levine, Stix, \&
  Harvey}]{altschuler77}
Altschuler, M.~D., Levine, R.~H., Stix, M., \& Harvey, J. 1977, \solphys, 51,
  345

\bibitem[{Arge {et~al.}(2004)Arge, Luhmann, Odstrcil, Schrijver, \&
  Li}]{arge04}
Arge, C., Luhmann, J., Odstrcil, D., Schrijver, C., \& Li, Y. 2004, Journal of
  Atmospheric and Solar-Terrestrial Physics, 66, 1295

\bibitem[{{Aschwanden} {et~al.}(2000a){Aschwanden}, {Alexander}, {Hurlburt},
  {Newmark}, {Neupert}, {Klimchuk}, \& {Gary}}]{aschwanden00a}
{Aschwanden}, M.~J., {Alexander}, D., {Hurlburt}, N., {Newmark}, J.~S.,
  {Neupert}, W.~M., {Klimchuk}, J.~A., \& {Gary}, G.~A. 2000, \apj,
  531, 1129

\bibitem[{Aschwanden \& Nitta(2000b)}]{aschwanden00}
Aschwanden, M.~J. \& Nitta, N. 2000, \apjl, 535, L59

\bibitem[{Aschwanden \& Acton(2001)}]{aschwanden01}
Aschwanden, M.~J. \& Acton, L.~W. 2001, \apj, 550, 475

\bibitem[{Aschwanden \& Schrijver(2002)}]{aschwanden02}
Aschwanden, M.~J. \& Schrijver, C.~J. 2002, \apjs, 142, 269

\bibitem[{Aschwanden(2008)}]{aschwanden08:asp}
Aschwanden, M.~J. 2008, in Astronomical Society of the Pacific Conference
  Series, Vol. 383, Subsurface and Atmospheric Influences on Solar Activity,
  ed. R.~Howe, R.~W. Komm, K.~S. Balasubramaniam, \& G.~J.~D. Petrie, 1

\bibitem[{Bercik {et~al.}(2005)Bercik, Fisher, Johns-Krull, {}, \&
  Abbett}]{bercik05}
Bercik, D.~J., Fisher, G.~H., Johns-Krull, C.~M., {}, \& Abbett, W.~P. 2005,
  \apj, 631, 529

\bibitem[{Cohen {et~al.}(2007)Cohen, Sokolov, Roussev, Arge, Manchester,
  Gombosi, Frazin, Park, Butala, Kamalabadi, \& Velli}]{cohen07}
Cohen, O., Sokolov, I.~V., Roussev, I.~I., Arge, C.~N., Manchester, W.~B.,
  Gombosi, T.~I., Frazin, R.~A., Park, H., {et al.} 2007, \apjl, 654, L163

\bibitem[{Cohen {et~al.}(2008)Cohen, Sokolov, Roussev, \& Gombosi}]{cohen08}
Cohen, O., Sokolov, I.~V., Roussev, I.~I., \& Gombosi, T.~I. 2008, \jgr, 113, A03104

\bibitem[{{Cohen} {et~al.}(2009){Cohen}, {Attrill}, {Manchester}, \&
  {Wills-Davey}}]{cohen09}
{Cohen}, O., {Attrill}, G. D.~R., {Manchester}, W.~B., \& {Wills-Davey}, M.~J.
  2009, \apj, 705, 587

\bibitem[{Delaboudini\`{e}re {et~al.}(1995)Delaboudini\`{e}re, Artzner,
  Brunaud, Gabriel, Hochedez, Millier, Song, Au, Dere, Howard, Kreplin,
  Michels, Moses, Defise, Jamar, Rochus, Chauvineau, Marioge, Catura, Lemen,
  Shing, Stern, Gurman, Neupert, Maucherat, Clette, Cugnon, \&
  Dessel}]{delaboudiniere95}
Delaboudini\`{e}re, Artzner, G.~E., Brunaud, J., Gabriel, A.~H., Hochedez,
  J.~F., Millier, F., Song, X.~Y., Au, B., {et al.} 1995, \solphys, 162, 291

\bibitem[{Frazin {et~al.}(2009)Frazin, Jacob, Manchester, Morgan, \&
  Wakin}]{frazin09}
Frazin, R.~A., Jacob, M., Manchester, W.~B., Morgan, H., \& Wakin, M.~B. 2009,
  \apj, 695, 636

\bibitem[{Freeland \& Handy(1998)}]{freeland98}
Freeland, S.~L. \& Handy, B.~N. 1998, \solphys, 182, 497

\bibitem[{Groth {et~al.}(2000)Groth, De~Zeeuw, Gombosi, \& Powell}]{groth00}
Groth, C. P.~T., De~Zeeuw, D.~L., Gombosi, T.~I., \& Powell, K.~G. 2000,
  \jgr, 105, 25,053

\bibitem[{{Jacobs} {et~al.}(2009){Jacobs}, {Roussev}, {Lugaz}, \&
  {Poedts}}]{jacobs09}
{Jacobs}, C., {Roussev}, I.~I., {Lugaz}, N., \& {Poedts}, S. 2009, \apjl, 695, L171

\bibitem[{{Landi} {et~al.}(2002){Landi}, {Feldman}, \& {Dere}}]{landi02}
{Landi}, E., {Feldman}, U., \& {Dere}, K.~P. 2002, \apjs, 139,
  281

\bibitem[{Landi {et~al.}(2006)Landi, Zanna, Young, Dere, Mason, \&
  Landini}]{landi06}
Landi, E., Zanna, D.~G., Young, P.~R., Dere, K.~P., Mason, H.~E., \& Landini,
  M. 2006, \apjs, 162, 261

\bibitem[{Lionello {et~al.}(2001)Lionello, Linker, {}, \& Mikic}]{lionello01}
Lionello, R., Linker, J.~A., {}, \& Mikic, Z. 2001, \apj,
  546, 542

\bibitem[{Lionello {et~al.}(2009)Lionello, Linker, \& Mikic}]{lionello09}
Lionello, R., Linker, J.~A., \& Mikic, Z. 2009, \apj, 690,
  902

\bibitem[{Lugaz {et~al.}(2007)Lugaz, Manchester, Roussev, T\'{o}th, \&
  Gombosi}]{lugaz07}
Lugaz, N., Manchester, Roussev, I.~I., T\'{o}th, G., \& Gombosi, T.~I. 2007,
  \apj, 659, 788

\bibitem[{Lugaz {et~al.}(2008)Lugaz, Vourlidas, Roussev, Jacobs, Manchester, \&
  Cohen}]{Lugaz08}
Lugaz, N., Vourlidas, A., Roussev, I.~I., Jacobs, C., Manchester, \& Cohen, O.
  2008, \apjl, 684, L111

\bibitem[{Lugaz {et~al.}(2009)Lugaz, Vourlidas, Roussev, \& Morgan}]{Lugaz09}
Lugaz, N., Vourlidas, A., Roussev, I., \& Morgan, H. 2009, \solphys, 259,
  269

\bibitem[{{Miki{\'{c}}} {et~al.}(1999){Miki{\'{c}}}, {Linker}, {Schnack},
  {Lionello}, \& {Tarditi}}]{mikic99}
{Miki{\'{c}}}, Z., {Linker}, J.~A., {Schnack}, D.~D., {Lionello}, R., \&
  {Tarditi}, A. 1999, Physics of Plasmas, 6, 2217

\bibitem[{Mok {et~al.}(2005)Mok, Mikic, Lionello, {}, \& Linker}]{mok05}
Mok, Y., Mikic, Z., Lionello, R., {}, \& Linker, J.~A. 2005, \apj, 621, 1098

\bibitem[{Mok {et~al.}(2008)Mok, Miki{\'c}, Lionello, \& Linker}]{mok08}
Mok, Y., Miki{\'c}, Z., Lionello, R., \& Linker, J.~A. 2008, \apjl, 679, L161

\bibitem[{Moses {et~al.}(1997)Moses, Clette, Delaboudini{\`ere}, Artzner,
  Bougnet, Brunaud, Carabetian, Gabriel, Hochedez, Millier, Song, Au, Dere,
  Howard, Kreplin, Michels, Defise, Jamar, Rochus, Chauvineau, Marioge, Catura,
  Lemen, Shing, Stern, Gurman, Neupert, Newmark, Thompson, Maucherat,
  Portier-Fozzani, Berghmans, Cugnon, van Dessel, \& Gabryl}]{moses97}
Moses, D., Clette, F., Delaboudini{\`ere}, J.~P., Artzner, G.~E., Bougnet, M.,
  Brunaud, J., Carabetian, C., Gabriel, A.~H., {et al.} 1997, \solphys, 175, 571

\bibitem[{Oran {et~al.}(2009)Oran, Sokolov, Roussev, Van Der~Holst, Manchester,
  \& Gombosi}]{oran09}
Oran, R., Sokolov, I.~V., Roussev, I.~I., Van Der~Holst, B., Manchester, W.~B.,
  \& Gombosi, T.~I. 2009, in Astronomical Society of the Pacific Conference
  Series, Vol. 406, Numerical Modeling of Space Plasma Flows: Astronum-2009
  Proceedings of the 2nd International Conference, ed. {N.~V.~Pogorelov,
  E.~Audit, P.~Colella, \& G.~P.~Zank}

\bibitem[{Parker(1988)}]{parker88}
Parker, E.~N. 1988, \apj, 330, 474

\bibitem[{Powell(1999)}]{powell99}
Powell, K. 1999, J. Comput. Phys., 154, 284

\bibitem[{Roussev {et~al.}(2003)Roussev, Gombosi, Sokolov, Velli, Manchester,
  DeZeeuw, Liewer, T\'{o}th, \& Luhmann}]{roussev03b}
Roussev, I.~I., Gombosi, T.~I., Sokolov, I.~V., Velli, M., Manchester, W.,
  DeZeeuw, D.~L., Liewer, P., T\'{o}th, G., \& Luhmann, J. 2003,
  \apjl, 595, L57

\bibitem[{{Roussev} {et~al.}(2004){Roussev}, {Sokolov}, {Forbes}, {Gombosi},
  {Lee}, \& {Sakai}}]{roussev04}
{Roussev}, I.~I., {Sokolov}, I.~V., {Forbes}, T.~G., {Gombosi}, T.~I., {Lee},
  M.~A., \& {Sakai}, J.~I. 2004, \apjl, 605, L73

\bibitem[{Roussev \& Sokolov(2006)}]{roussev06}
Roussev, I.~I. \& Sokolov, I.~V. 2006, Washington DC American Geophysical Union
  Geophysical Monograph Series, 165, 89

\bibitem[{Roussev {et~al.}(2007)Roussev, Lugaz, \& Sokolov}]{roussev07}
Roussev, I.~I., Lugaz, N., \& Sokolov, I.~V. 2007, \apjl., 668,
  L87

\bibitem[{Schrijver {et~al.}(2004)Schrijver, Sandman, Aschwanden, \&
  Derosa}]{schrijver04}
Schrijver, C.~J., Sandman, A.~W., Aschwanden, M.~J., \& Derosa, M.~L. 2004,
  \apj, 615, 512

\bibitem[{Sokolov {et~al.}(2009)Sokolov, Roussev, Skender, Gombosi, \&
  Usmanov}]{sokolov09}
Sokolov, I.~V., Roussev, I.~I., Skender, M., Gombosi, T.~I., \& Usmanov, A.~V.
  2009, \apj, 696, 261

\bibitem[{{Spitzer}(1965)}]{spitzer65}
{Spitzer}, L. 1965, {Physics of fully ionized gases}, { 2nd rev. ed.} edn.
  ({New York}: {Interscience Publication})

\bibitem[{T\'{o}th {et~al.}(2005)T\'{o}th, Sokolov, Gombosi, Chesney, Clauer,
  De~Zeeuw, Hansen, Kane, Manchester, Oehmke, Powell, Ridley, Roussev, Stout,
  Volberg, Wolf, Sazykin, Chan, Yu, \& K\~{a}³ta}]{toth05}
T\'{o}th, G., Sokolov, I.~V., Gombosi, T.~I., Chesney, D.~R., Clauer, R.~C.,
  De~Zeeuw, D.~L., Hansen, K.~C., Kane, K.~J., {et al.} 2005, \jgr, 110, A12226

\bibitem[{Tsuneta {et~al.}(1991)Tsuneta, Acton, Bruner, Lemen, Brown,
  Caravalho, Catura, Freeland, Jurcevich, \& Owens}]{tsuneta91}
Tsuneta, S., Acton, L., Bruner, M., Lemen, J., Brown, W., Caravalho, R.,
  Catura, R., Freeland, {et al.} 1991, \solphys, 136,
  37

\bibitem[{Withbroe(1988)}]{withbroe88}
Withbroe, G.~L. 1988, \apj, 325, 442

\end{thebibliography}
\end{document}